\documentclass[twocolumn]{aastex631}
\usepackage{amsmath}
\usepackage{graphicx}   
\usepackage{bm}         
\usepackage{natbib}
\usepackage{etoolbox}
\usepackage{rotating}
\usepackage{array}
\usepackage{booktabs}
\usepackage{amssymb}
\usepackage{multirow}

\newcommand{\sm}{${M}_\odot$}
\newcommand{\sL}{${L}_\odot$}
\newcommand{\tbol}{$T_{\rm bol}$}
\newcommand{\lbol}{$L_{\rm bol}$}
\newcommand{\iras}{IRAS 15398$-$3359}

\newcommand{\hhco}{H$_{2}$CO}

\newcommand{\co}{C$^{18}$O}
\newcommand{\ccchh}{c-C$_3$H$_2$}

\newcommand{\meta}{CH$_3$OH}

\newcommand{\hthcop}{H$^{13}$CO$^+$}

\newcommand{\kms}{km s$^{-1}$}
\newcommand{\mjybeam}{mJy beam$^{-1}$}
\newcommand{\feii}{[Fe~\textsc{ii}]}
\newcommand{\neii}{[Ne~\textsc{ii}]}
\newcommand{\si}{[S~\textsc{i}]}
\newcommand{\niii}{[Ni~\textsc{ii}]}
\newcommand{\av}{A$_\text{V}$}
\begin{document}

\title{CORINOS. III. Outflow Shocked Regions of the Low-mass Protostellar Source IRAS 15398-3359 with JWST and ALMA}

\author[0000-0003-3655-5270]{Yuki Okoda}
\affiliation{NRC Herzberg Astronomy and Astrophysics, 5071 West Saanich Road, Victoria, BC, V9E 2E7, Canada}
\affiliation{Star and Planet Formation Laboratory, RIKEN Cluster for Pioneering Research, 2-1, Hirosawa, Wako-shi, Saitama 351-0198, Japan}

\author[0000-0001-8227-2816]{Yao-Lun Yang}
\affiliation{Star and Planet Formation Laboratory, RIKEN Cluster for Pioneering Research, 2-1, Hirosawa, Wako-shi, Saitama 351-0198, Japan}

\author[0000-0001-5175-1777]{Neal J. Evans II}
\affiliation{Department of Astronomy, The University of Texas at Austin, Austin, TX 78712, USA}

\author[0000-0001-8064-2801]{Jaeyeong Kim}
\affiliation{Korea Astronomy and Space Science Institute, 776 Daedeok-daero, Yuseong-gu Daejeon 34055, Republic of Korea}

\author[0000-0002-4801-436X]{Mihwa Jin}
\affiliation{Astrochemistry Laboratory, Code 691, NASA Goddard Space Flight Center, Greenbelt, MD 20771, USA}
\affiliation{Department of Physics, Catholic University of America, Washington, DC 20064, USA}

\author[0000-0001-7723-8955]{Robin T. Garrod}
\affiliation{Departments of Chemistry and Astronomy, University of Virginia, Charlottesville, VA 22904, USA}

\author[0000-0001-8822-6327]{Logan Francis}
\affiliation{Leiden Observatory, Leiden University, PO Box 9513, 2300 RA Leiden, The Netherlands}

\author[0000-0002-6773-459X]{Doug Johnstone}
\affiliation{NRC Herzberg Astronomy and Astrophysics, 5071 West Saanich Road, Victoria, BC, V9E 2E7, Canada}
\affiliation{Department of Physics and Astronomy, University of Victoria, Victoria, BC, V8P 5C2, Canada}

\author[0000-0001-9664-6292]{Cecilia Ceccarelli}
\affiliation{Univ. Grenoble Alpes, CNRS, IPAG, 38000 Grenoble, France}

\author[0000-0003-1514-3074]{Claudio Codella}
\affiliation{INAF, Osservatorio Astrofisico di Arcetri, Largo E. Fermi 5, I-50125, Firenze, Italy}
\affiliation{Univ. Grenoble Alpes, CNRS, IPAG, 38000 Grenoble, France}

\author[0000-0002-7570-5596]{Claire J. Chandler}
\affiliation{National Radio Astronomy Observatory, PO Box O, Socorro, NM 87801, USA}

\author[0000-0002-9865-0970]{Satoshi Yamamoto}
\affiliation{The Graduate University for Advanced Studies SOKENDAI, Shonan Village, Hayama, Kanagawa 240-0193, Japan}
\affiliation{Research Center for the Early Universe, The University of Tokyo, 7-3-1, Hongo, Bunkyo-ku, Tokyo 113-0033, Japan}

\author[0000-0002-3297-4497]{Nami Sakai}
\affiliation{Star and Planet Formation Laboratory, RIKEN Cluster for Pioneering Research, 2-1, Hirosawa, Wako-shi, Saitama 351-0198, Japan}


\begin{abstract}

While molecular outflows have been studied in details with radio interferometry, observations of the hotter gas in protostellar outflows at a comparable physical scale is often challenging. Combined with ALMA, JWST allows us to investigate the cold and hot gas with unprecedented spatial resolution and sensitivity. We present a detailed comparison between the gas distributions probed with ALMA and JWST in the primary outflow of IRAS 15398$-$3359. At 2000 au scale, the southwestern outflow shows four shell structures in 5--10 micron continuum, whereas the submillimeter H$_2$CO emission traces two of the four shells closest to the protostar. Submillimeter emission from CS, CCH, c-C$_3$H$_2$, and CH$_3$OH shows the same two shells, and the $^{12}$CO emission covers most of the outflow region. SO and SiO only trace a condensation at the edge of the shell closest to the protostar. None of these lines observed with ALMA show the outermost shell. At 500 au scale, we find hot H$_2$ gas inside the outflow cavity with JWST. The derived temperature of H$_2$ is 1147$\pm$198 K within a 0\farcs5 aperture at the protostar. The foreground mass column density of dust is (1.4--2.0)$\times$10$^{-3}$ g$\cdot$cm$^{-2}$ (\av=47--66 mag) in the outflow, using the dust model from \citet{2001ApJ...548..296W}. We also find an 8$^{\circ}$ difference between the directions toward the [Fe II] knot and the outermost shell in the MIRI image, which may be interpreted as the precession of the [Fe II] jet. The dynamical timescale of the [Fe II] knot is 10 yrs, suggesting a current event.


\end{abstract}

\section{Introduction}\label{sec:intro}
\par Understanding the formation processes of young low-mass protostellar systems is of fundamental importance for exploring the diversity of later planetary systems and the origin of the solar system.
In the earliest phases, outflows driven by magnetohydrodynamical processes remove the angular momentum of the infalling gas to further enable gas accretion onto protostars \citep[e.g.,][]{2016ARA&A..54..491B, 2019ApJ...876..149M}.
Hence, outflows are deeply connected to stellar mass assembly, disk formation, and, through feedback, the ongoing infall in the envelope \citep[e.g.,][]{2020MNRAS.491.2180M, 2020ApJ...896..158T, 2023ASPC..534..317T}.

Observational studies at millimeter/submillimeter wavelengths have identified components within the outflow structure, such as cavity walls and shock-heated regions, using molecular lines \citep[e.g.,][]{2014ApJ...795..152O, 2016AA...587A.145B, 2020A&A...633A.126B, 2021A&A...655A..65T, 2022ApJ...927...54O,2024AJ....167...72D}.
Furthermore, pure rotational lines of molecular hydrogen (H$_2$) at mid-infrared wavelengths are one of the common tracers for directly probing the shocked gas within outflows \citep[e.g.,][]{1998ApJ...506L..75N, 2003ApJ...590L..41L, 2009ApJ...698.1244M}.
Recently, H$_2$ lines have been observed in protostellar outflows with JWST \citep{2024A&A...685C...5G, 2024ApJ...962L..16N, 2024A&A...687A..36T}.
Using rotation diagram analysis, these lines reveal that the outflows have warm ($\sim$500 K) and hot ($\sim$1000 K) components, providing new insight into the protostellar system structure at the few 100 au scale.


\par \iras\ is a solitary low-mass protostellar source at the Class 0 phase \citep[\tbol$=$44 K, \lbol$=$1.8 \sL;][]{2013ApJ...779L..22J} located in the Lupus I molecular cloud \citep[$d=$154.9$\pm$3.4 pc;][]{2020AA...643A.148G}.
An outflow along the northeast to southwest direction has been detected with ALMA \citep{2014ApJ...795..152O, 2017ApJ...834..178Y, 2020ApJ...900...40O, 2021AA...648A..41V}, SMA \citep{2016AA...587A.145B}, and JCMT/APEX \citep{2015AA...576A.109Y}.
The outflow direction has been reported to be at a P.A. of 215\degr-230\degr\ (measured north through east) based on the ALMA and SMA observations, and 
the inclination angle has been evaluated to be 20 \degr \citep{2014ApJ...795..152O}, which is close to an edge-on (0 \degr) configuration.
\cite{2016AA...587A.145B} and \cite{2021AA...648A..41V} suggest precession of the primary outflow, following the distribution of $^{12}$CO emission. 
\cite{2021AA...648A..41V} determine four bipolar structures which vary along a clockwise direction.
Emission from \hhco, \meta, SO, CS, CCH, and \ccchh\ trace the cavity walls and/or shock-heated regions in the outflow \citep{2014ApJ...795..152O, 2018ApJ...864L..25O, 2020ApJ...900...40O}.
The dynamical timescale of the outflow has been reported to be $\sim$10$^2$-10$^3$ yr \citep{2014ApJ...795..152O, 2015AA...576A.109Y, 2016AA...587A.145B, 2021AA...648A..41V}, indicating the youth of this protostar.
Interestingly, a relic outflow along the northwest to southeast direction (P.A.=140\degr) was detected in emission from \hhco, \co, SO, SiO, and \meta\ \citep{2021ApJ...910...11O}.
As well, extended outflow features were additionally found along the north-south direction \citep[P.A. = 0\degr;][]{2024ApJ...966..192S}.  
These outflow directions are significantly different from the younger outflow, directly connected to the protostellar system.
\cite{2021ApJ...910...11O} suggest that the outflow direction might be changing as the protostar evolves.
 
A disk structure has already formed around the protostar \citep{2018ApJ...864L..25O, 2023ApJ...958...60T}.
These authors analyzed the SO emission in the vicinity of the protostar.
\cite{2018ApJ...864L..25O} reported a protostellar mass of 0.007$^{+0.004}_{-0.003}$ \sm\ by fitting a Keplerian rotation at a resolution of $\sim$0\farcs2.  
\cite{2023ApJ...958...60T} conducted a twice higher-resolution observation, and reported a protostellar mass to be 0.022-0.1 \sm\ based on a power-law fit to the SO emission. 
The disk mass was reported to be between 0.006 \sm\ and 0.001 \sm\ from 1.2 mm dust continuum emission, assuming dust temperatures between 20 K and 100 K \citep{2018ApJ...864L..25O}.
Meanwhile, the envelope mass \citep[0.5-1.2 \sm:][]{2012AA...542A...8K, 2013ApJ...779L..22J} is significantly greater than the estimated star and disk masses.
Combined, these results support the early evolutionary stage of this protostar.

\par Recently, observations with JWST-MIRI were conducted toward this source as part of the CORINOS (COMs ORigin Investigated by the Next-generation
Observatory in Space) program \citep{2022ApJ...941L..13Y, 2024ApJ...974...97S}.  \cite{2022ApJ...941L..13Y} found four shell-like structures in the southwestern outflow as well as rich molecular hydrogen and atomic ion lines in the outflow near the protostar.
In this paper, we study the morphology and physical parameters of the \iras\ primary outflow along the northeast to southwest direction, using both JWST and ALMA.
JWST, together with ALMA, facilitate a comprehensive understanding of the physical structure of young protostars.
We describe the observations in Section \ref{sec:obs}. 
In Section \ref{sec:Morphology}, we update and newly report on images observed with JWST, discussing the findings by comparing to the molecular emission imaged with ALMA.
The method for extinction correction and the derivation of H$_2$ gas physical parameters are summarized in Section \ref{sec:TempH2}.
In Section \ref{sec:precession}, we describe the observed difference between the direction of the jet and the outflow, before concluding the paper in Section \ref{sec:summary}.


\section{Observations}\label{sec:obs}
\subsection{JWST}
The JWST images were taken by the Mid-Infrared Instrument (MIRI; \citealt{2015PASP..127..584R}; \citealt{2015PASP..127..612B};  \citealt{2023PASP..135d8003W}) as part of the CORINOS program \citep{2022ApJ...941L..13Y}.  The observations were carried out as simultaneous MIRI imaging while performing spectroscopy in medium-resolution spectrometer (MRS) mode.  The observations were performed on 2022, July 20 with two fields of view (FoVs), corresponding to the science and background pointings of the MRS observations \citep{2022ApJ...941L..13Y}.  The center of the imager, which has a FoV of 74\arcsec$\times$113\arcsec, is separated by $\sim$1.5\arcmin\ from the center of the MRS FoV. The MRS FoV is $\sim$6\arcsec$\times $$\sim$6\arcsec (See also Figures \ref{atomic} and \ref{fig:h2wex}). The background pointing of the MRS observation covers images of the blue-shifted (southwestern) outflow of IRAS 15398$-$3359, which is one of the main observations presented in this study.  

For each MRS pointing (science and background), MIRI images were taken with the three F560W, F770W, and F1000W filters centered at 5.6, 7.7, and 10.0 \micron, respectively.  The observation details are provided in Table\,\ref{tbl:miri_imaging_params}.  The imaging data were reduced with JWST pipeline v.1.14.0 \citep{2023zndo..10022973B} using the calibration reference data \texttt{jwst\_1231.pmap}. The JWST data reduction pipeline has three stages focusing on detector-level calibration, instrument-level and observing-mode correction, and combination of all exposures, respectively. Two sets of the MIRI images, corresponding to two MRS pointings, were reduced separately from raw data to Stage 2.  In Stage 3, two imaging observations for each filter were reduced together, producing an image with almost twice the FoV.  The spectroscopic data were reduced following the steps described by \citet{2022ApJ...941L..13Y} using the JWST pipeline v.1.12.5 and calibration reference data \texttt{jwst\_1183.pmap}.

\par 
The line parameters are summarized in Table \ref{tbl:line_jwst_params}.  The line emission maps are obtained by performing Gaussian fitting at each pixel, taking into account an underlying linear continuum.  After running line fitting over all pixels, we integrate the continuum-subtracted intensity over the determined spectral full width at half maximum (FWHM).
The line maps provided in this paper show the intensities higher than the 3$\sigma$ noise level, where $\sigma$ is the standard deviation of the residual after the line fitting.

\begin{table}[htbp!]
    \centering
    \caption{Observation Parameters of the MIRI Imaging}
    \label{tbl:miri_imaging_params}
    \begin{tabular}{ccccc}
        \toprule
         Filter & Exposure time (s) & $\lambda_{0}^{a}$ & $\Delta\lambda$ & FWHM \\
          & (s) & (\micron) & (\micron) & (\arcsec) \\
         \midrule
         F560W  & 1410 & 5.6  & 1.2 & 0.207 \\
         F770W  & 3541 & 7.7  & 2.2 & 0.269 \\
         F1000W & 1388 & 10.0 & 2.0 & 0.328 \\
         \bottomrule
         \multicolumn{5}{l}{$^{a}$The central wavelength of the filter.} \\
    \end{tabular}
\end{table}

\begin{table}[htbp!]
    \centering
    \caption{Line Parameters Observed with JWST}
    \label{tbl:line_jwst_params}
    \begin{tabular}{ccc}
        \toprule
         Species & Transition & $\lambda_{0}$ (\micron)$^{a}$ \\
         \midrule
         H$_2$  &  S(1) & 17.03484\\
                &  S(2) &12.27861\\
                &  S(3) &9.66492\\
                &  S(4) &8.02505\\
                &  S(5) &6.90952\\
                &  S(6) &6.10857\\
                &  S(7) &5.51118\\
                &  S(8) &5.05312\\
         \feii  & $^6F_{7/2}$-$^6F_{9/2}$ &25.988290\\
         \feii  & $^4F_{7/2}$-$^4F_{9/2}$ &17.935950\\
         \feii  & $^4F_{9/2}$-$^6F_{9/2}$ &5.3401690\\
         \neii  & $^2P_{1/2}$-$^2P_{3/2}$ &12.813550\\
         \si    & $^3P_{1}$-$^3P_{2}$ &25.249000\\
         \niii  & $^2D_{3/2}$-$^2D_{5/2}$ & 6.636000 \\
         \bottomrule
         \multicolumn{3}{l}{$^{a}$The central wavelength of the line.} 
         \\
    \end{tabular}
\end{table}

\subsection{ALMA}\label{sec:ALMAobs}

\begin{table*}[htbp!]

  \caption{Line Parameters Observed with ALMA$^a$ \label{ALMA_observations}}
  \centering
  \begin{tabular}{cccccc}
  \hline
  Line  & Transition & Frequency  &  Beam size  & \multicolumn{2}{c}{Velocity range$^b$} \\
        &            & (GHz)      &             &  \multicolumn{2}{c}{(start, end (\kms))} \\
  \hline
  $^{12}$CO & $J=2-1$                             & 230.5380000           & 0\farcs64$\times$0\farcs56 (P.A. = 76.5\degr)     & ($-$9.0, 4.8) & (6.0, 7.2) \\
  \hhco     & ($J, K_a, K_c$) = 3$_{0,3}-2_{0,2}$                   & 218.2221920           & 0\farcs36$\times$0\farcs30 (P.A. = 63.1\degr)     & (2.0, 5.0)    & (5.2, 7.2) \\
  CS        & $J=5-$4                             & 244.9355565           & 0\farcs19$\times$0\farcs16 (P.A. = $-$86.6\degr)  & (0.5, 5.14)   & (5.3, 7.2) \\
  CCH       & $N = 3-2$, $J = 7/2-5/2$, $F = 4-3$ & 262.0042600           & \multirow{4}{*}{0\farcs18$\times$0\farcs15 (P.A. = 88.3\degr)}     & \multirow{4}{*}{(3.2, 5.2)}   & \multirow{4}{*}{(5.3, 7.2)} \\
                & $N = 3-2$, $J = 7/2-5/2$, $F = 3-2$ & 262.0064820           &  &  \\
                & $N = 3-2$, $J = 5/2-3/2$, $F = 3-2$ & 262.0649860           &  &  \\
                & $N = 3-2$, $J = 5/2-3/2$, $F = 2-1$ & 262.0674690           &  &  \\
  \multirow{2}{*}{\ccchh}   & ($J, K_a, K_c$) = 6$_{0,6} - 5_{1,5}$  &\multirow{2}{*}{217.8221480}   & \multirow{2}{*}{0\farcs36$\times$0\farcs30 (P.A. = 62.8\degr)}     & \multirow{2}{*}{(4.4, 5.0)}    & \multirow{2}{*}{(5.2, 6.0)} \\
  & ($J, K_a, K_c$) = $6_{1,6} - 5_{0,5}$&  &     &     &  \\
  \meta     & ($J, K_a, K_c$) = 4$_{2,3}-3_{1,2}$                   & 218.4400630           & 0\farcs36$\times$0\farcs30 (P.A. = 62.8\degr)     & \multicolumn{2}{c}{(2.8, 5.6)} \\
  SiO      & $J=5-$4                             & 217.1049190           & 0\farcs37$\times$0\farcs30 (P.A. = 62.4\degr)     & \multicolumn{2}{c}{(1.6, 4.6)} \\
  SO        & $J_N=6_5-5_4$                         & 219.9494420           & 0\farcs36$\times$0\farcs30 (P.A. = 58.7\degr)     & \multicolumn{2}{c}{(1.6, 4.6)} \\
  \hline
  \end{tabular}
  \begin{flushleft}
  \tablecomments{
  $^a$ The $^{12}$CO line data are taken from the ALMA archival data \citep[2013.1.00879.S;][]{2013ApJ...772...22Y}, and the other line data are observed in the ALMA Large Program FAUST \citep[2018.1.01205.L;][]{2021FrASS...8..227C}.
  $^b$Integrated velocity ranges for Figure \ref{moment_alma}. The left and right ranges of $^{12}$CO, \hhco, CS, CCH, and \ccchh\ are for the blueshifted and redshifted components, respectively.
  }
  \end{flushleft}
\end{table*}

\par 
The molecular lines from ALMA observations used in our analysis are listed with their parameters in Table \ref{ALMA_observations}.
The \hhco, \meta, SiO, SO, CS, CCH, and \ccchh\ lines were observed in Band 6 as part of the ALMA Large Program FAUST \citep[Fifty AU STudy of the chemistry in the disk/envelope system of solar-like protostars\footnote{http://faust-alma.riken.jp}:][2018.1.01205.L]{2021FrASS...8..227C}.
Single-field observations for two frequency setups from 232 to 235 GHz and from 246 to 248 GHz were carried out between 2018 October and 2019 January.
The field center was taken to be ($\alpha_{2000}$, $\delta_{2000}$)= (15$^{\rm h}$43$^{\rm m}$02$^{\rm s}$.242, $-$34\arcdeg 09\arcmin 06\farcs805) for both setups. 
In this study, we used the 12\,m array data from two array configurations (C43-5 and C43-2 for sparse and compact configurations, respectively) and the 7\,m array data, combining these visibility data sets in the UV plane. 
Further observation parameters are described by \citet{2021ApJ...910...11O, 2023ApJ...948..127O}.  We also use the ALMA archival data for $^{12}$CO (2013.1.00879.S) in this paper. The details for this specific observation are described by \cite{2017ApJ...834..178Y}.

\par Higher-resolution observations for the \hhco\ (3$_{0,3}-2_{0,2}$, 3$_{2,1}-2_{2,0}$, and 3$_{2,2}-2_{2,1}$) lines were taken at Band 6 on 2021 July 18, 20, 21, 26, and 27 in the ALMA program of 2019.1.01359.S.
The lines and observation parameters including calibrator sources are summarized in Tables \ref{ALMA_observations_2019} and \ref{parameter_observation}, respectively.
We combined the four execution blocks in the $uv$ plane.
The field center was taken to be ($\alpha_{2000}$, $\delta_{2000}$)= (15$^{\rm h}$43$^{\rm m}$01$^{\rm s}$.316, $-$34\arcdeg09\arcmin15\farcs300).
The continuum peak position of \iras\ is ($\alpha_{2000}$, $\delta_{2000}$) =  (15$^{\rm h}$43$^{\rm m}$02$^{\rm s}$.232, $-$34\arcdeg09\arcmin06\farcs958), which is within the field of view of 21\arcsec.
The primary beam (half-power beam) width is 26\farcs5.

\par These data were reduced in Common Astronomy Software Applications (CASA) package 6.4.1 \citep{2007ASPC..376..127M}.
Both phase and amplitude self-calibration were carried out for the FAUST observation (2018.1.01205.L) using line-free continuum emission for each configuration, where the details are described by \cite{2022ApJ...934...70I}.  
Spectral cubes were prepared through the procedure of CLEANing the dirty images with a Briggs robustness parameter of 0.5.
In these observations, the absolute flux calibration uncertainty is 10\%.

\section{Morphology of the Outflow}\label{sec:Morphology}
The MIRI images (F560W, F770W, and F1000W) and the spatial distribution of H$_2$ S(5), H$_2$ S(7), \feii\ (25 $\micron$ and 17 $\micron$), \neii, and \si\ lines have been reported previously by \citet{2022ApJ...941L..13Y}.
We utilize these images and newly present emission maps of H$_2$ from S(1) to S(8) corrected for extinction, \feii\ (5.34 $\micron$), and \niii\ (6.64 $\micron$). As well, an RGB MIRI image is constructed to reveal the large scale outflow morphology.
We introduce the MIRI images in Section \ref{sec:results_miri} and the spatial distributions of H$_2$ and ions in Sections \ref{sec:h2} and \ref{sec:ions}, respectively.
In order to understand the outflow morphology comprehensively, we compare both the large scale ($\sim$2000 au) MIRI image of F560W and the small scale ($\sim$500 au) H$_2$ S(1) image against the molecular line distributions imaged with ALMA in Sections \ref{sec:comparison_large} and \ref{sec:comparison_small}, respectively.
For comparison, we focus on the $^{12}$CO, \hhco, \meta, SiO, SO, CS, CCH, and \ccchh\ line emission detected with ALMA.

\begin{figure*}[htbp!]
    \centering
    \includegraphics[scale=0.4]{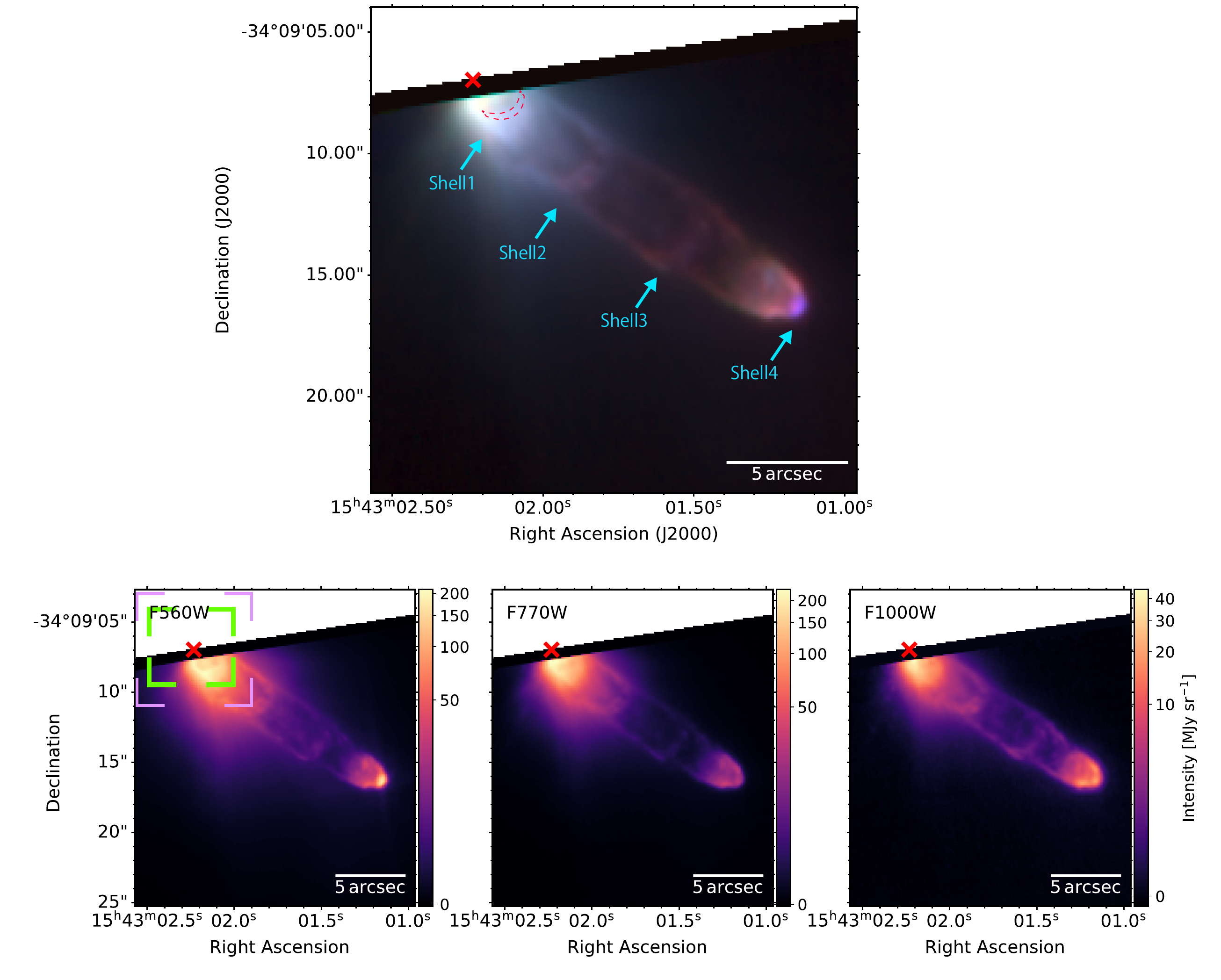}
    \caption{MIRI images. RGB is in the upper and the individual filters in the bottom. MIRI RGB image built from a composite of the three filter (F560W, F770W, and F1000W) MIRI images. In the upper panel, the area surrounded by the red dashed lines shows the first shell structure. In the bottom-left panel, the region with green lines correspond to that of H$_2$ in Figure \ref{h2_line}. The region with pink lines corresponds to those shown in Figures \ref{atomic} and \ref{h2_alma}.
    The red cross marks the continuum peak ($\alpha_{2000}$, $\delta_{2000}$)= (15$^{\rm h}$43$^{\rm m}$02$^{\rm s}$.232, $-$34\arcdeg09\arcmin06\farcs971) as determined by the ALMA observation (2021.1.00357.S). \label{miri}}
\end{figure*}

\subsection{Large Scale Morphology}\label{sec:comparison_large_all}
\subsubsection{Broadband Images}\label{sec:results_miri}
\par Figure\,\ref{miri} shows the MIRI image at each filter as well as the newly combined RGB image.
As reported by \cite{2022ApJ...941L..13Y}, these images reveal four shell-like structures and a terminal knot in the southwestern blue-shifted outflow, which is most clearly seen in the F560W image. 
Theoretical studies suggest potential mechanisms for the shell-like structures in \iras. 
Such shell-like features could imply an alternative origin of outflow substructures, for example creation by MHD-driven shocks \citep{2023ApJ...944..230S}.
Alternatively, Hanawa et al. (in prep). find that these substructures can arise naturally from an underexpanded jet.
Also, an episodic ejection could produce these features. The primary outflow of \iras\ observed in $^{12}$CO by \cite{2021AA...648A..41V} has been suggested to eject episodically, although the shell-like structures are not seen clearly with their images.
Moreover, the previous \hthcop\ observation suggests an accretion burst during the last 100-1000 yr, which is similar to the dynamical timescale of the primary outflow (See Section \ref{sec:precession}).

In addition to the emission from the collimated outflow, there is extended midinfrared emission around the continuum peak (F560W in Figure \ref{miri}), which seems to have a wide opening angle of the outflow, likely dominated by scattered light. This feature is most prominent in the F560W image. 
The northeastern outflow and the protostar are outside the FoV of the imager.
The RGB image reveals a blue color for the tip of the outermost shell (Shell 4 in Figure \ref{miri}), suggesting a relatively high temperature. As put forward by \cite{2022ApJ...941L..13Y}, the feature appears similar to a bow-shock. In general, we assume that thermal emission and scattered light dominates these images; however, each of these filters include H$_2$ lines, and hence, the H$_2$ emission could affect the intensity as well. 
To separate the H$_2$ lines from continuum, spectral mapping using the IFU will be necessary, since MIRI does not have narrow band filters.

\subsubsection{Comparison between MIRI Continuum (F560W) and ALMA Molecular Lines}\label{sec:comparison_large}
\par 
Figure \ref{moment_alma} shows the ALMA observerd moment 0 maps of $^{12}$CO, \hhco, CS, CCH, and \ccchh, integrated over blue-shifted and red-shifted velocities.
The blue-shifted components of $^{12}$CO extend to the southwest direction from the protostar, covering the main features seen in the MIRI images except for the outermost shell.
While most of the red-shifted $^{12}$CO emission appears in the northeastern outflow, we also see red-shifted $^{12}$CO emission inside the blue-shifted outflow between the third shell and the outermost shell (Figure \ref{moment_alma}(a)).
The first and second shells also appear to emit in the \hhco, CS, and CCH lines.
The \ccchh\ emission traces the first shell, and its intensity is higher than 3$\sigma$ at the edge of the second shell ($\sigma$ = 1 \mjybeam$\cdot$\kms).
The \hhco\ and CS lines also show faint emission, at about the 2$\sigma$ level ($\sigma$ = 3 \mjybeam$\cdot$\kms), at the third shell position in the red-shifted component of the southwest outflow.
The molecular lines observed by ALMA therefore reveal the collimated outflow, though the MIRI images appear to trace a slightly wider opening angle as well.
Note that additional extended structure is revealed in the maps of \hhco, CS, and \ccchh\ along the northwest to southeast direction around the protostar. This emission is located at the position of the relic outflow previously studied by \cite{2021ApJ...910...11O}.

\begin{figure*}[htbp!]
    \centering
    \includegraphics[scale=0.4]{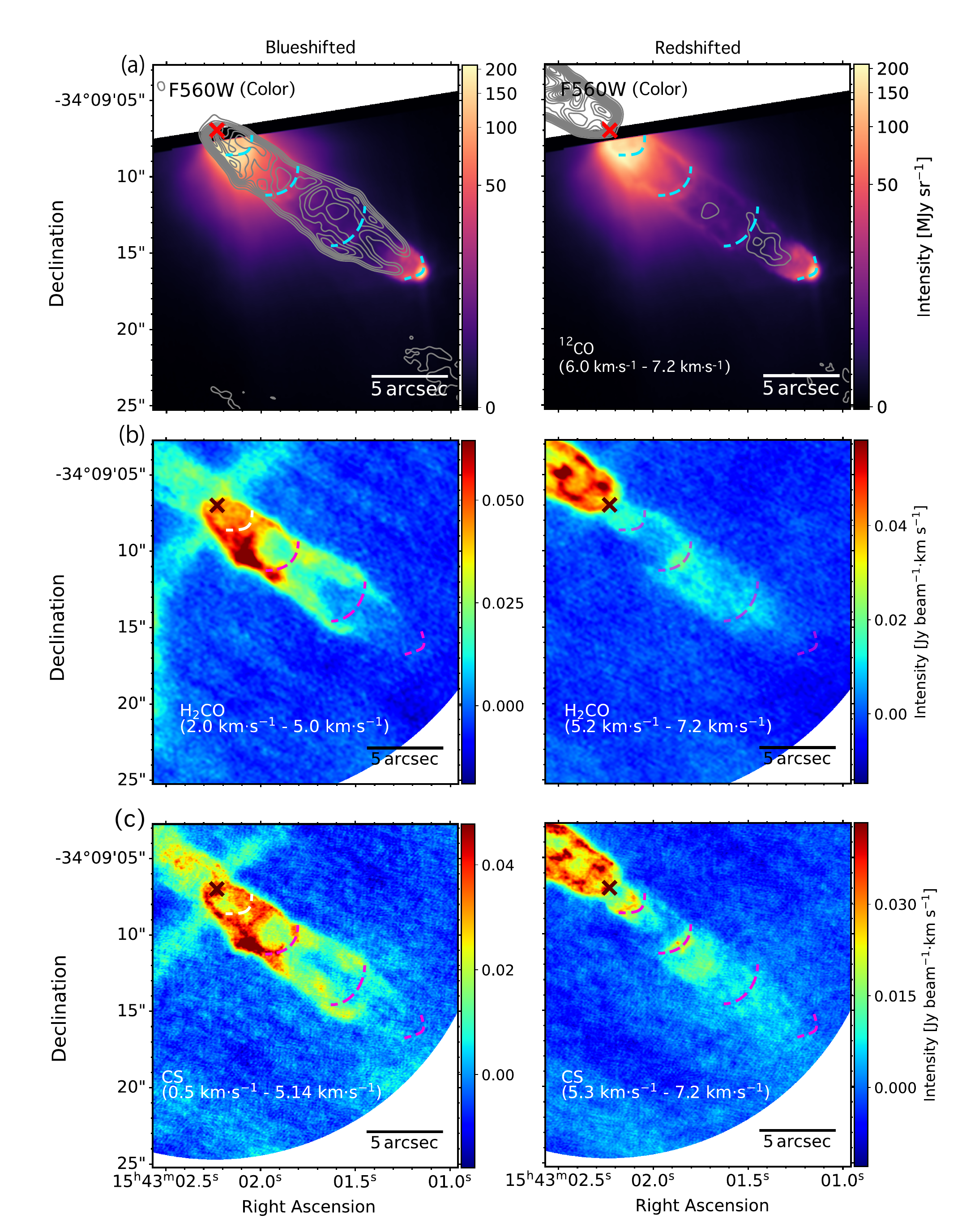}
    \caption{(a) Moment 0 maps of the $^{12}$CO line emission (contours) overlaid on the MIRI F560W image.  The blue-shifted and red-shifted CO components are shown in the left and right panels, respectively.  Contour levels are plotted for every 3$\sigma$ from 3$\sigma$, where $\sigma$ is 78 and 11 \mjybeam channel$^{-1}$ for the blue- and red-shifted components, respectively.  (b)--(e) The moment 0 maps of the blue-shifted (left) and red-shifted (right) components of \hhco\ (3$_{0,3}-2_{0,2}$), CS ($J=$5$-$4), CCH ($N = 3-2$, $J = 7/2-5/2$, $F = 4-3$ and 3$-$2, and $N = 3-2$, $J = 7/2-5/2$, $F = 4-3$ and 2$-$1), and \ccchh\ (6$_{0,6}-5_{1,5}$ and 6$_{1,6}-5_{0,5}$). The ``x'' marks the continuum peak measured from the ALMA observation, which is 
    ($\alpha_{2000}$, $\delta_{2000}$) = 15$^{\rm h}$43$^{\rm m}$02$^{\rm s}$.232, $-$34\arcdeg09\arcmin06\farcs971.  The velocities on each panel show the integrated velocity ranges.  The four shell-like structures identified in the MIRI images are indicated as cyan, magenta, and white dashed lines.
    \label{moment_alma}}
\end{figure*}

\addtocounter{figure}{-1}
\begin{figure*}[htbp!]
    \centering
    \includegraphics[scale=0.4]{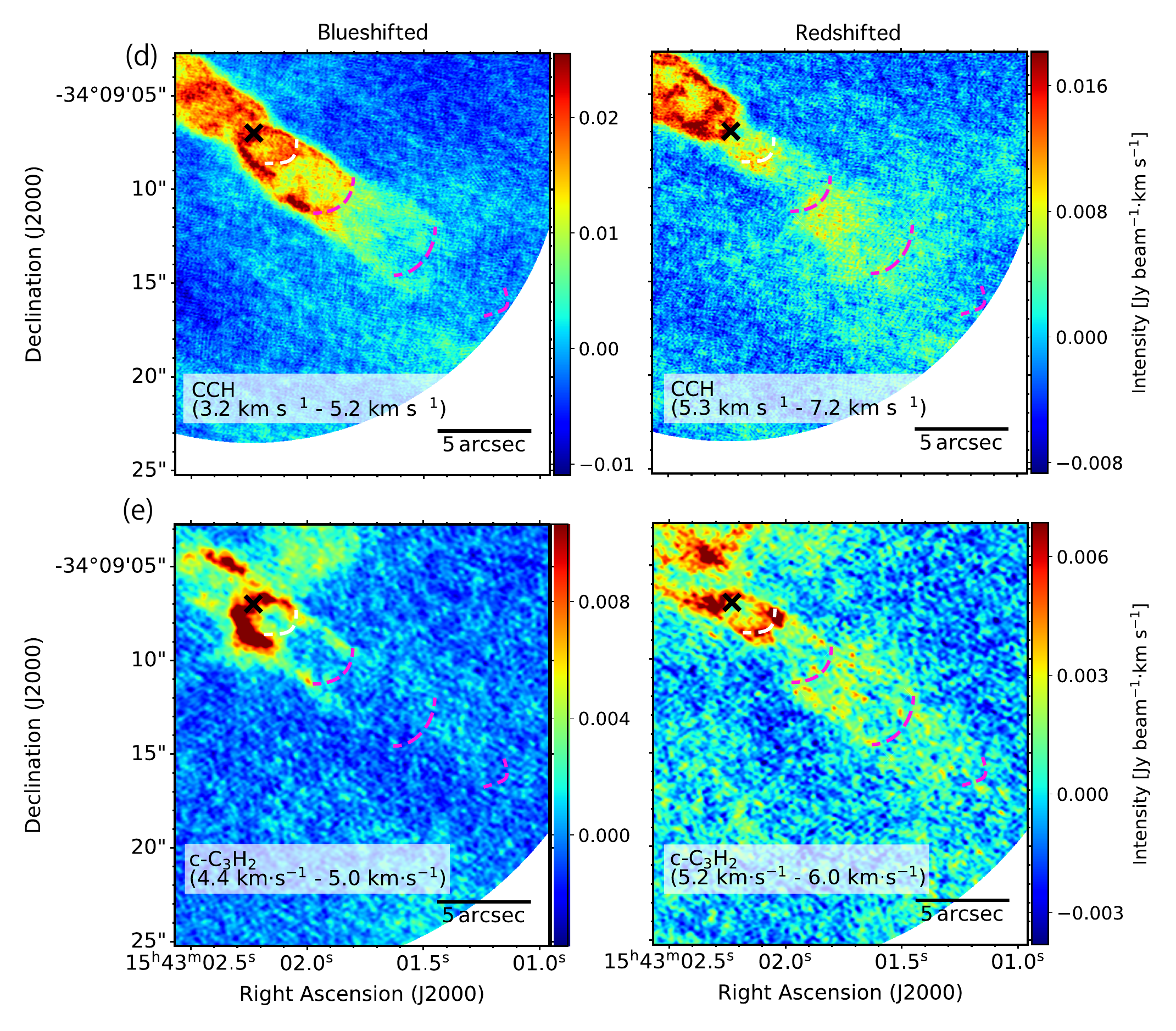}
    \caption{(Continued)}
\end{figure*}

\begin{figure*}[htbp!]
    \centering
    \includegraphics[width=\textwidth]{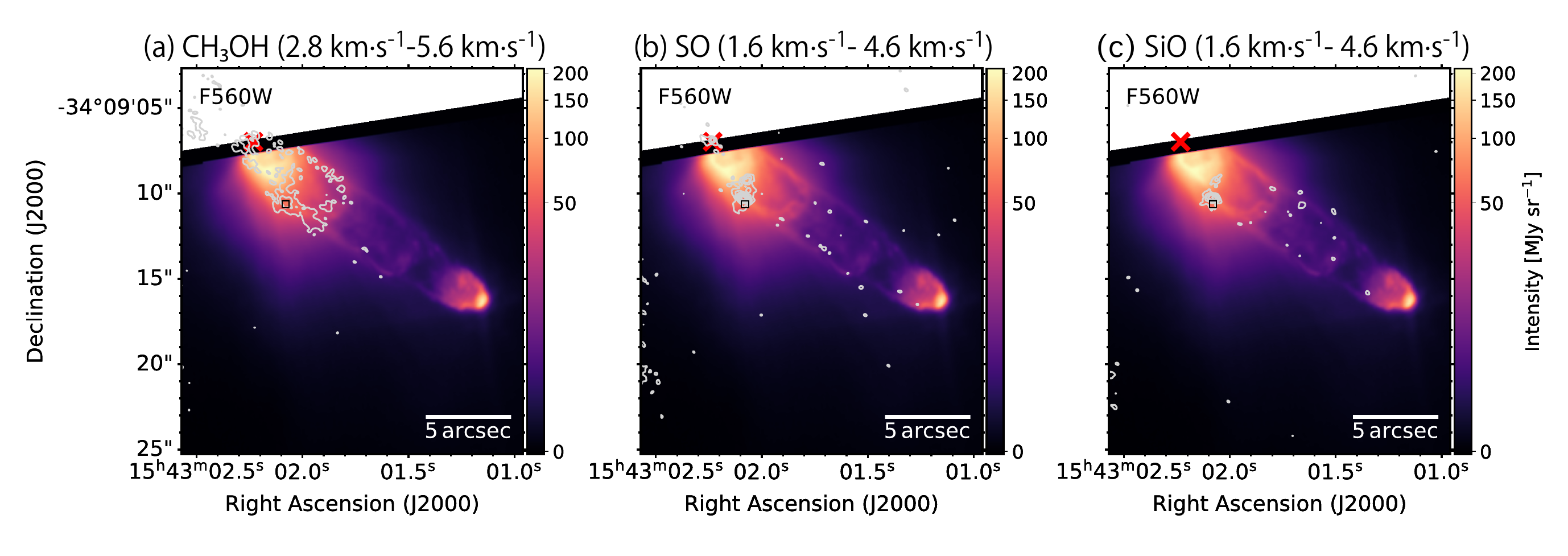}
    \caption{(a, b, c) Moment 0 maps of \meta, SO, and SiO line emission overlaid on the MIRI F560W image. Contour levels for \meta, SiO, and SO are every 3$\sigma$ from 3$\sigma$, where $\sigma$ is 2, 2, and 2.5 \mjybeam channel$^{-1}$, respectively. The ``x'' marks the continuum peak measured from the ALMA observation, which is ($\alpha_{2000}$, $\delta_{2000}$) = 15$^{\rm h}$43$^{\rm m}$02$^{\rm s}$.232, $-$34\arcdeg09\arcmin06\farcs971. The black square represents Blob D, located at ($\alpha_{2000}$, $\delta_{2000}$) = 15$^{\rm h}$43$^{\rm m}$02$^{\rm s}$.08, $-$34\arcdeg09\arcmin10\farcs63, as reported by \cite{2020ApJ...900...40O}. The velocities on each panel show the integrated velocity ranges. \label{moment_shocks}}
\end{figure*}


Figure \ref{moment_shocks} shows the integrated intensity maps of \meta, SO, and SiO observed with ALMA overlaid on the MIRI image at F560W. These are well known key molecules to understand the shock-heated gas \citep[e.g.,][]{1992ApJ...392L..87M, 1997ApJ...487L..93B}, which should be released by sputtering from the dust mantles into the gas-phase.
The \meta\ emission traces the first and second shells in addition to the outflow cavity wall. 
The SiO and SO emission only traces a compact structure in the outflow close to the southeastern part of the first shell, also seen in the \meta\ emission, where the velocity widths are found to be 1.8 \kms.
The SO and SiO compact structures are thought to be a shock-heated region, which is consistent with Blob D suggested by \citet{2020ApJ...900...40O}. 
We highlight that jet-like structure is seen in neither the high-velocity components of the ALMA observations presented in this study nor any previous ALMA observations \cite[e.g,][]{2014ApJ...795..152O, 2021AA...648A..41V}.
Since the jet is mostly atomic or ionized, it would likely be escaped from  ALMA and previous radio observations. Furthermore, previous IR observations might not have had enough resolution and sensitivity to detect the jet. 

The outermost shell in the MIRI image is not seen clearly in any of the molecular lines observed with these ALMA observations, nor in previous studies with low-resolution submillimeter observations.
It is located toward the edge of the ALMA field of view, and hence the sensitivity is not high enough to detect the outermost shell.
Even with the observation of 2019.1.01359.S, where the outermost shell is fully covered, the \hhco\ emission is not apparent (Figure \ref{spectra_hhco}). It may, however, be resolved-out because of the high resolution observation (Tables \ref{ALMA_observations_2019} and \ref{parameter_observation}). This shell may have a high temperature, as suggested by its blue color in the RGB image (Figure \ref{miri}). It would therefore be expected to have much weaker emission from the lower excitation molecular lines observable by ALMA.

\begin{figure*}[htbp!]
  \centering
  \includegraphics[width=\textwidth]{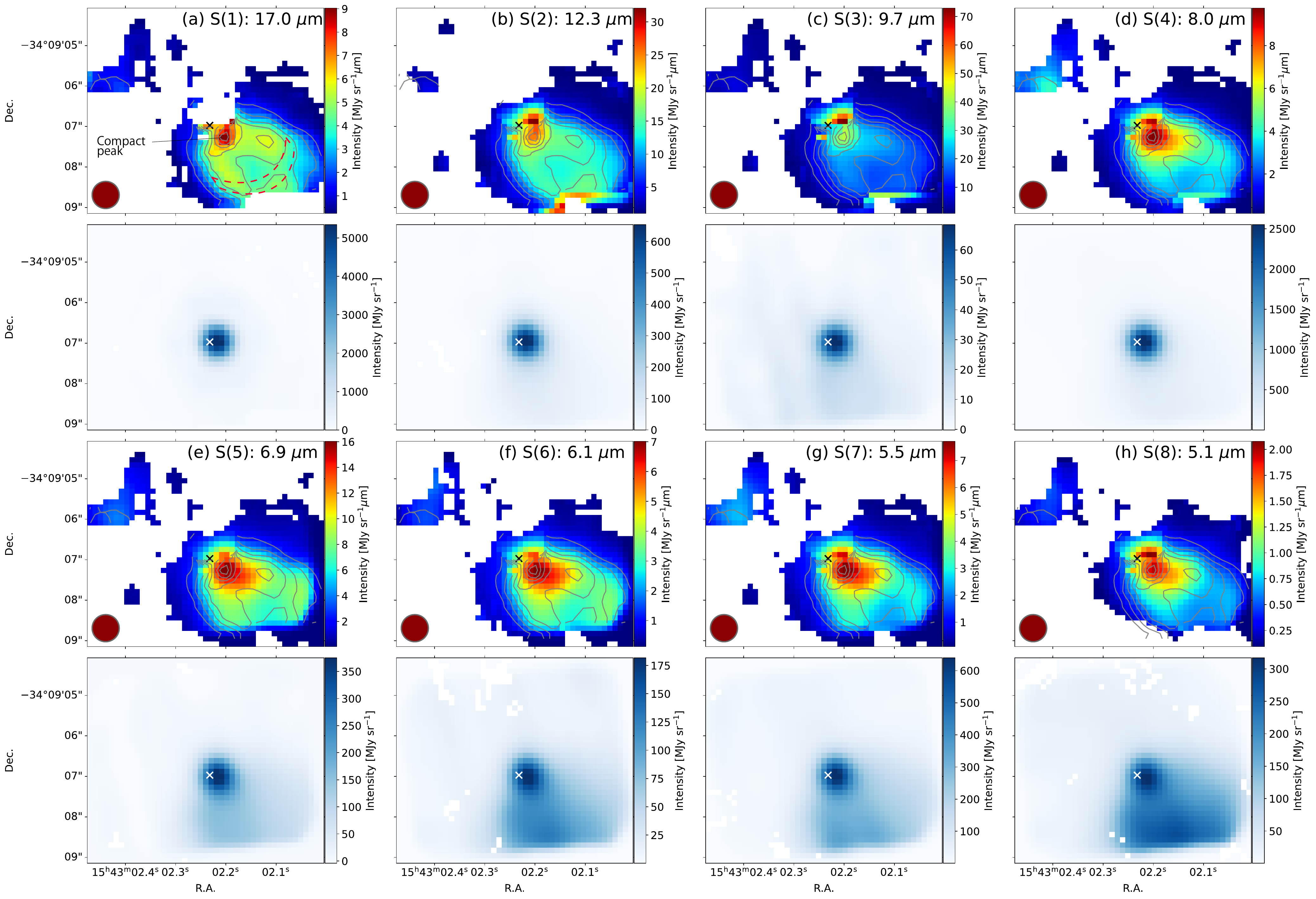}
  \caption{Integrated intensity maps for the H$_2$ lines after extinction correction (upper panels) and the continuum emission at the corresponding wavelength (lower panels). The contours of the S(1) map are overlaid on each top panel. The contour levels are every 2$\sigma$ from 3$\sigma$, where $\sigma$ is 0.5 MJy sr$^{-1}$ \micron. 
  The ``x'' marks the continuum peak position ($\alpha_{2000}$, $\delta_{2000}$)= (15$^{\rm h}$43$^{\rm m}$02$^{\rm s}$.232, $-$34\arcdeg09\arcmin06\farcs971) measured from the ALMA data (2021.1.00357.S). The area with red dashed lines in the top of panel (a) shows the position of the first shell structure in the MIRI image (Figure \ref{miri}). The black circle represents the beam size of 0\farcs67.
  \label{h2_line}}
\end{figure*}

\subsection{Small Scale}
\label{sec:comparison_small_all}
The JWST MIRI MRS observations measured several molecular and ionized emission lines at $\sim3\arcsec-5\arcsec$ around the protostar, probing outflow and jet gas at a smaller scale compared to that presented in Section\,\ref{sec:comparison_large_all}. We show the results of H$_2$ and ionized emission in separate sections below.
For the H$_2$ lines, we use extinction corrected maps, as we study the physical parameters in Section \ref{sec:lte-H2}.
The methods for dealing with the extinction correction are deferred to Section \ref{sec:extinction}.
\cite{2022ApJ...941L..13Y} reported the detection of pure rotational lines from S(1) to S(8) of H$_2$ in the outflow and the distributions of H$_2$ S(5) and H$_2$ S(7) without extinction correction, which we also present in Figure\,\ref{fig:h2wex}.  

\subsubsection{Distribution of H$_2$ Emission}
\label{sec:h2}
\par 

Before the extinction correction (Figure\,\ref{fig:h2wex}), the observed H$_2$ maps show cavity walls in the lower excitation transitions from S(1) to S(3) in the southwestern outflow. In the higher excitation lines, two compact structures appear inside the cavity.
These two structures have a P.A. of 240$\pm$10\degr, closely aligned with the northeastern outflow \citep[P.A.=245\degr;][]{2021AA...648A..41V}.
The typical line width of the H$_2$ lines is 30 - 60 \kms, which is broader than the $^{12}$CO line width reported by \citet{2021AA...648A..41V}.
In the northeastern outflow, weaker extended emission is detected in the S(1), S(4), S(5), and S(7) lines. The S(4) line also shows a clearly compact structure to the northeast. 

The H$_2$ maps after correcting for extinction are shown in the top of each panel of Figure \ref{h2_line}. The region of the figure corresponds to that surrounded by green lines in the MIRI image of F560W (Figure\,\ref{miri}).
We re-gridded the spectral cubes to have the same pixel size of 0\farcs13, and smooth the spatial resolution to 0\farcs67 ($\theta_L$). 
This spatial resolution is based on the wavelength of the S(1) line at 17.03484 \micron\ ($\lambda_L$), using the equation reported by \citet{2023AJ....166...45L} as, 
\begin{equation}
  \theta_L = 0.033 (\lambda_L/\micron) + 0\farcs106. \label{eq:theta}
\end{equation}
After applying the extinction correction (Figure \ref{h2_line}), the intensity distributions remain qualitatively the same. The boundary of the observed H$_2$ distribution coincides with the shell structure closest to the protostar identified from the MIRI RGB image (Shell 1 in Figure\,\ref{miri}), which is the area with red dashed lines in Figure \ref{h2_line}(a).  Part of the cavity wall is seen in the S(1) line.
The compact peak is seen around $\sim0.5\arcsec$ from the protostar in the southwestern lobe in all of the lines. 
The emission in the high excitation lines of S(4) to S(8) is slightly extended toward the southwest from the compact peak, compared to that in the low excitation lines of S(1) to S(3).
This is similar to the two compact structures in Figure\,\ref{fig:h2wex} as mentioned above. 


The continuum from 5 to 17 \micron\ is measured as part of the line fitting process, as shown in the bottom of each panel of Figure \ref{h2_line}.  The protostar appears as a point source at longer wavelengths ($\geq 8 \micron$) and tends to become extended toward the southwestern outflow at shorter wavelengths ($< 8 \micron$).  Extended emission shows up at 9.7, 6.9, 6.1, 5.5, and 5.1 \micron. Interestingly, the extended continuum emission, likely due to scattering, 
has an alignment only slightly offset westward of south,
which is significantly different from the direction of the extended H$_2$ emission.  At the shortest wavelength (5.1 \micron), the extended emission starts collimated from the protostar and becomes wider at larger distance.  A similar structure is also seen in the continuum at 6.1 \micron.

\begin{figure*}[htbp!]
  \centering
  \includegraphics[width=\textwidth]{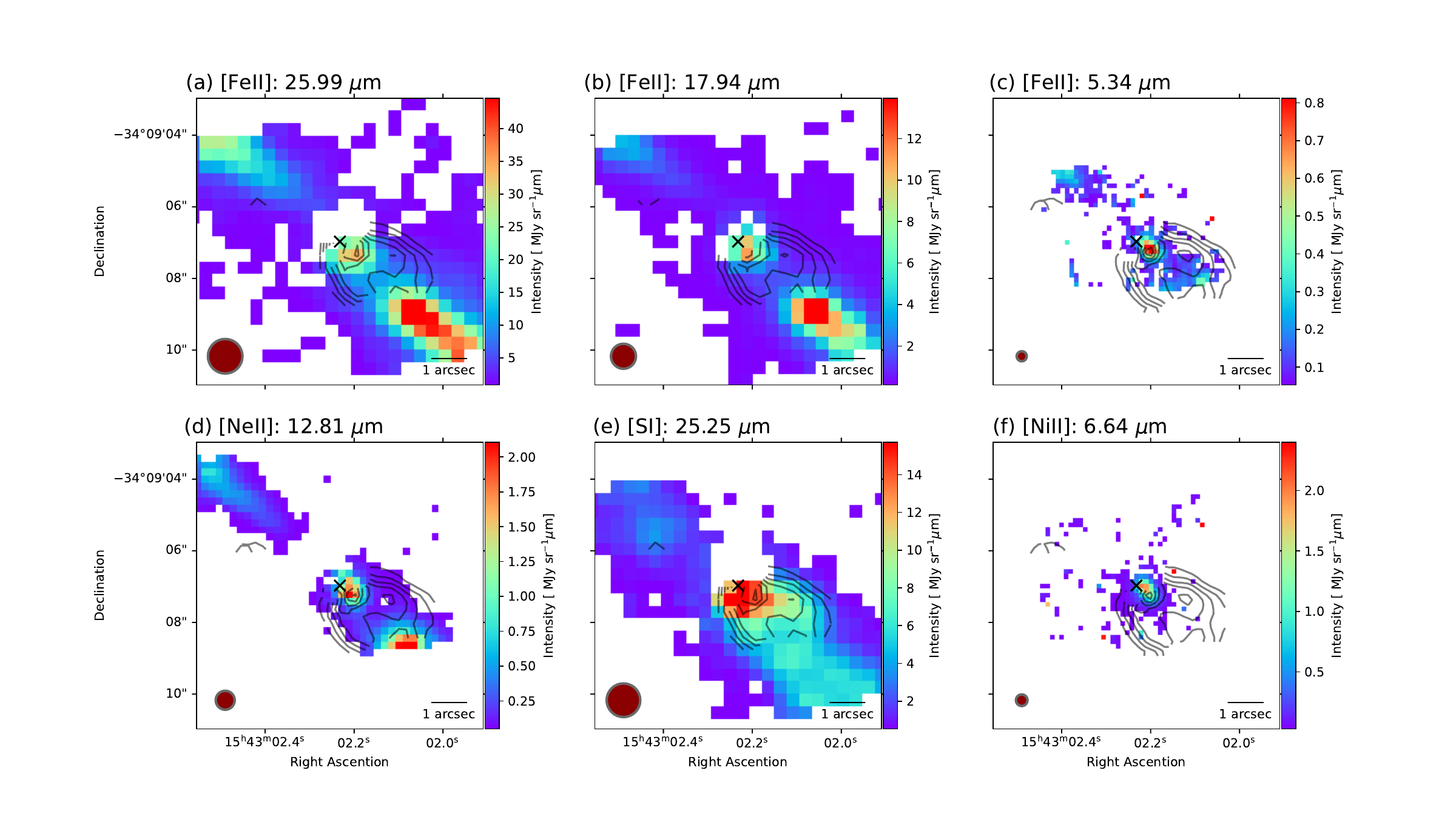}
  \caption{
  Integrated intensity maps of \feii\ (25 \micron, 17 \micron, and 5 \micron), \neii, \si, and \niii.
  Contours show the integrated intensity map of the H$_2$ S(1) line. The contour levels are every $\sigma$ from 3$\sigma$, where $\sigma$ is 0.5 MJy sr$^{-1}$ \micron. The contours look different from panel to panel because the H$_2$ S(1) emission is smoothed to match that of the comparing lines. The ``x'' marks the continuum peak position ($\alpha_{2000}$, $\delta_{2000}$) = (15$^{\rm h}$43$^{\rm m}$02$^{\rm s}$.232, $-$34\arcdeg09\arcmin06\farcs971) as determined by the ALMA observation (2021.1.00357.S). The black circle represents each beam size.\label{atomic}}
\end{figure*}

\subsubsection{\rm \feii, \neii, \si, and \niii}\label{sec:ions}
Figure \ref{atomic} shows the integrated intensity maps of \feii\ $^6F_{7/2}$-$^6F_{9/2}$ (25.99 \micron), $^4F_{7/2}$-$^4F_{9/2}$ (17.94 \micron), and $^4F_{9/2}$-$^6F_{9/2}$ (5.34 \micron), \neii\ $^2P^0_{1/2}$--$^2P^0_{3/2}$ (12.81 \micron), \si\ $^3P_1$--$^3P_2$ (25.25 \micron), and \niii\ $^2D_{3/2}$--$^2D_{5/2}$ (6.64 \micron). These maps are not corrected for extinction.
The maps of \feii\ (25 \micron\ and 17 \micron) and \neii\ show a collimated structure, and compared against the H$_2$ shown in contours (Figures \ref{atomic}(a), (b), and (d)).
Maps of these two lines previously have been shown and discussed by \cite{2022ApJ...941L..13Y}, suggesting a jet-like structure within the outflow from \iras.
In general, both \feii\ and \neii\ are common tracers of shocked gas \citep[e.g.,][]{2024ApJ...966...41F}.
In our case, this is further supported by the directional alignment of the two peaks in the southwestern outflow.
Such structures are also seen by JWST in other low-mass protostars \citep{ 2024ApJ...962L..16N, 2024A&A...687A..36T}.
The \feii\ (5.34 \micron) and \niii\ (6.64 \micron) are newly reported here.
Figure \ref{atomic}(c) shows the compact emission of \feii\ (5.34 \micron) at $\sim0.5\arcsec$ offset from the protostar in the outflow (P.A.=240\degr). We find that a similar peak can also be seen in the \niii\ emission (Figure \ref{atomic}(f)) as well as in the two \feii\ lines at longer wavelengths detected with lower resolution (Figures \ref{atomic}(a) and (b)).  This peak coincides with the compact emission in the H$_2$ line, and could trace the jet-launching point.  On the other hand, the \si\ emission only has a single extended component at each outflow lobe, compared to that of \feii\ and \neii.  As shown in Figure \ref{atomic}(e), the \si\ distribution is comparable to the maps of H$_2$, although it has a moderately smaller opening angle at larger distance from the protostar (Figure\,\ref{atomic}(d)).  Thus, the \si\ emission is possibly tracing a different component than the jet.

\subsubsection{Comparison between MIRI H$_2$ S(1) and ALMA Molecular Lines} \label{sec:comparison_small}
\begin{figure*}[htbp!]
    \centering
    \includegraphics[width=\textwidth]{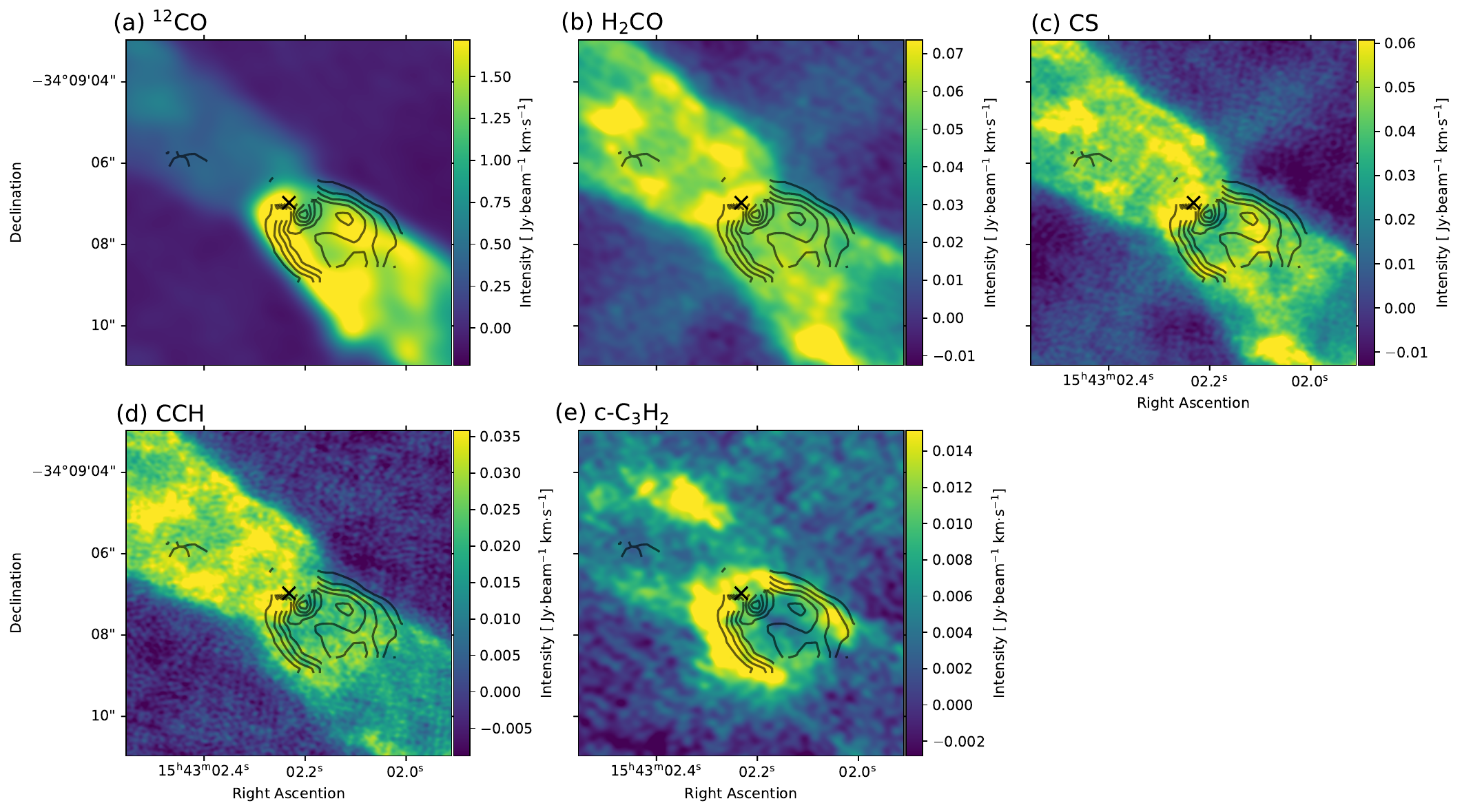}
    \caption{Integrated intensity maps of $^{12}$CO, \hhco, CS, CCH, and \ccchh\ (a--e), where the contours show the H$_2$ S(1) map as shown in the upper panel of Figure \ref{h2_line}(a). The ``x'' marks the continuum peak with ALMA observation, which is 
    ($\alpha_{2000}$, $\delta_{2000}$) = 15$^{\rm h}$43$^{\rm m}$02$^{\rm s}$.232, $-$34\arcdeg09\arcmin06\farcs971.  \label{h2_alma}}
\end{figure*}

Figure \ref{h2_alma} shows the moment 0 maps of the CO, \hhco, CS, CCH, and \ccchh\ lines near the protostar, where the H$_2$ S(1) map is shown in contours.  We here focus on these lines with ALMA for comparison at a smaller scale.
The outline observed with H$_2$ is quite consistent with that seen in these molecular lines with ALMA.
The \ccchh\ emission of the left outflow cavity in the southwest traces outside the H$_2$ S(1) emission, and the most collimated outflow among these five molecular lines is seen in the $^{12}$CO emission. 
The molecular line emission with ALMA is seen in the outflow cavity and the first shell structure, while the  H$_2$ S(1) emission to the southwest has a peak within 0\farcs5 of the protostar.
Thus, H$_2$ emission with JWST provides information about an additional small scale physical structure not seen by the molecular lines observed with ALMA.

\begin{figure}
    \centering
    \includegraphics[width=0.47\textwidth]{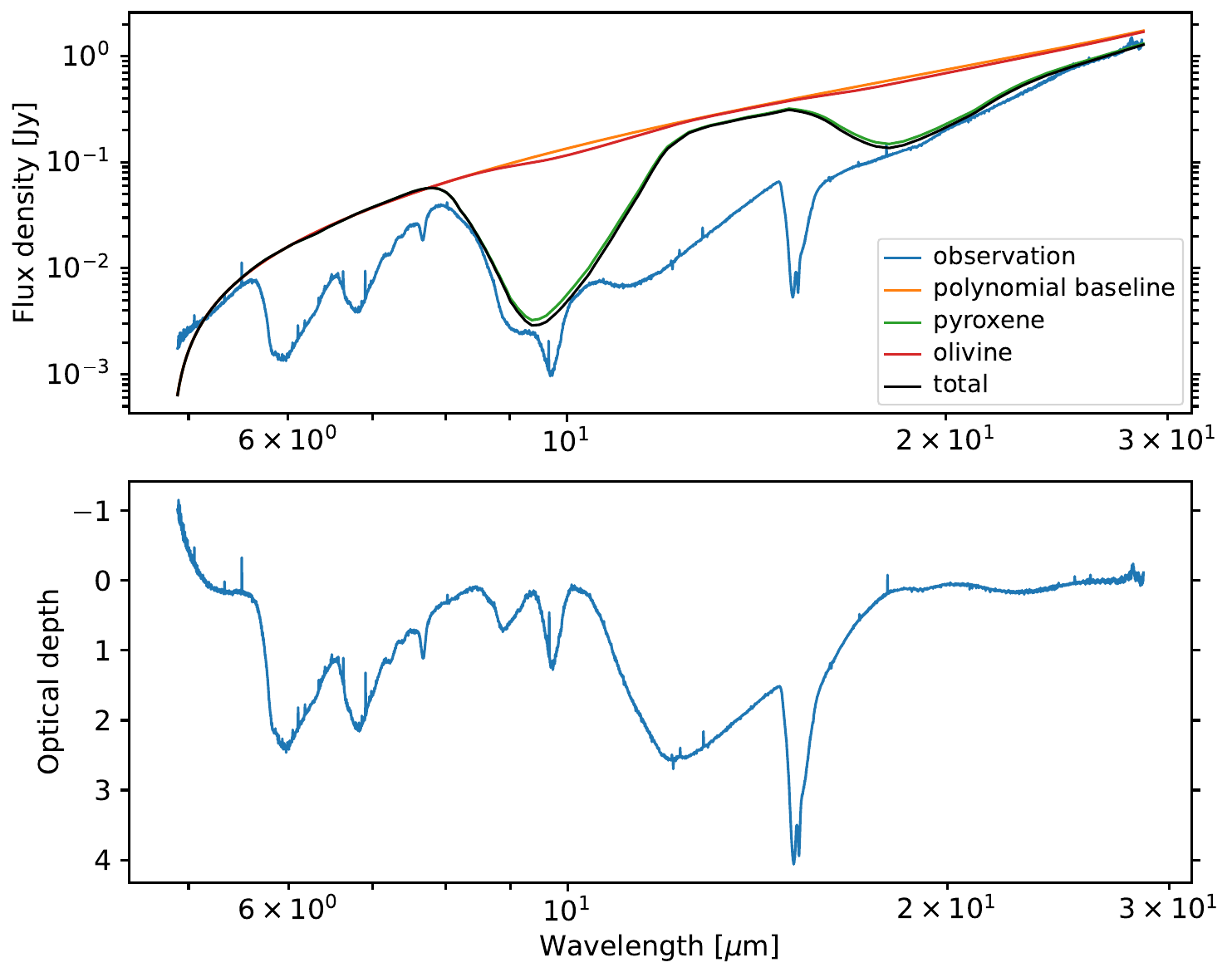}
    \caption{Top: MIRI MRS spectrum of \iras\ overlaid with the best-fitting synthetic silicate components of olivine and pyroxene. Bottom: The derived ice optical depth spectrum based on the fitted components showed in the top panel.}
    \label{fig:spec_silicate}
\end{figure}

\section{H$_2$ Temperature and Column Density}
\label{sec:TempH2}
\par We find that the region traced by H$_2$ emission likely has different physical conditions from that traced by molecular lines with ALMA (Section \ref{sec:comparison_small}). To understand physical parameters near the protostar, we derive the distributions of H$_2$ excitation temperature and column density by fitting the rotational diagram at each pixel, using the S(1) to S(8) lines.
Before the fitting, the H$_2$ lines need to be corrected for extinction due to dust grains and ice mantles. The visual extinction map is modeled as described in Section \ref{sec:extinction}.
The dust models used are summarized in Section \ref{sec:dust model}.
The equations for the LTE analysis are presented in Section \ref{sec:lte-H2}.
In the main text, we utilize a single-temperature fitting with an ice-free dust model, considering an additional ice absorption using the observed ice optical depth spectrum.
The results are discussed in Section \ref{sec:single-icefree} and are compared to those of another low-mass protostellar source, IRAS 16253--2429 reported by \cite{2024ApJ...962L..16N}, in Section \ref{sec:comparison_temp}.
For the analysis, we use the spectral cubes of H$_2$ with a pixel size of 0\farcs13 and a spatial resolution of 0\farcs67 (Figure \ref{h2_line}), as described in Section \ref{sec:h2}.


\begin{figure*}[htbp!]
    \centering
    \includegraphics[width=\textwidth]{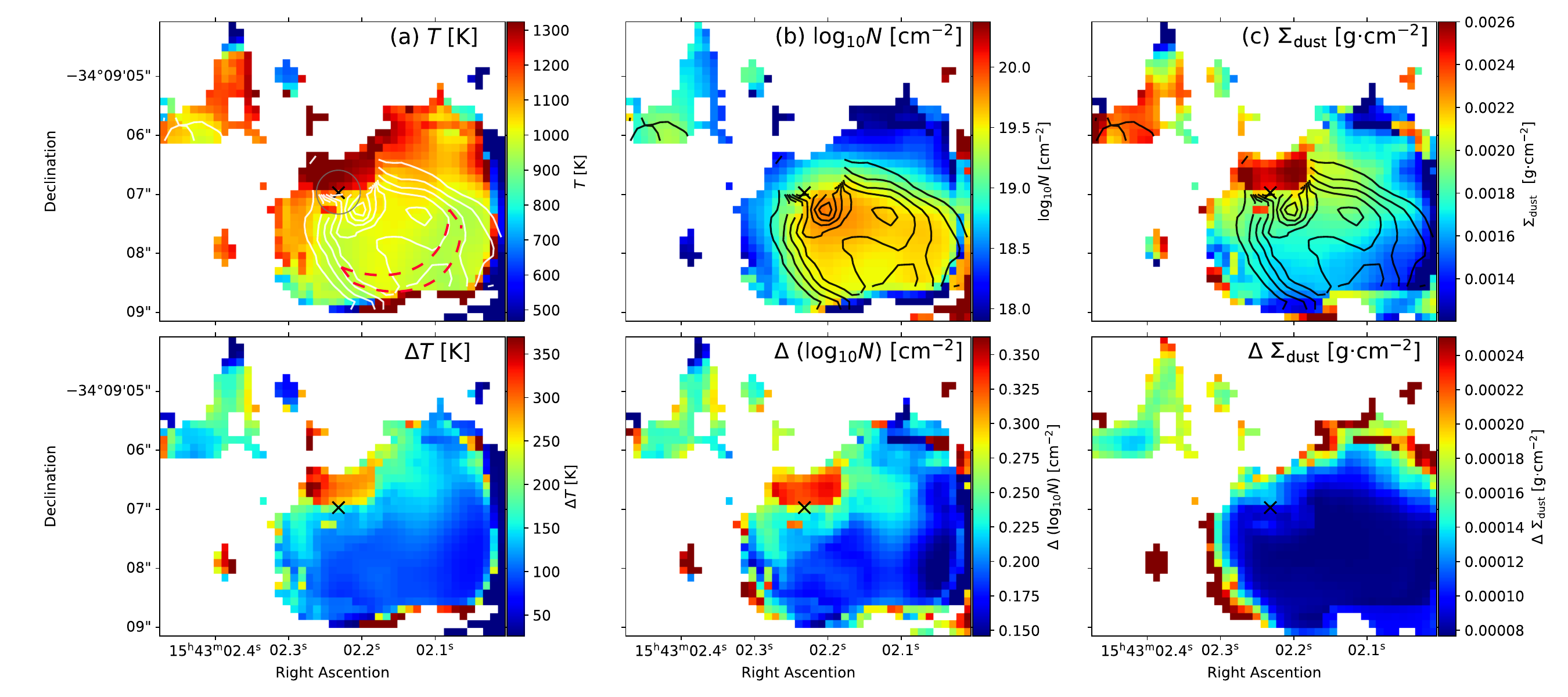}
    \caption{(a, b) The excitation temperature and column density maps of H$_2$ ($T$(H$_2$) and $N$(H$_2$), respectively). (c) Mass column density of dust, $\Sigma_{\rm dust}$. Bottom images of each panel show the uncertainties of the corresponding upper panel. We show the distributions of $T$(H$_2$) whose value is higher than three times the uncertainty. For those of $N$(H$_2$) and $\Sigma_{\rm dust}$, each value higher than each uncertainty is shown. These values are calculated for each pixel where four or more lines are detected. The ``x'' marks the protostar position ($\alpha_{2000}$, $\delta_{2000}$) = 15$^{\rm h}$43$^{\rm m}$02$^{\rm s}$.232, $-$34\arcdeg09\arcmin06\farcs971. Contours plot the H$_2$ S(1) map shown in Figure \ref{h2_line}. The area with red dashed lines in the top of panel (a) shows the position of the shell structure in the MIRI image (Figure \ref{miri}). The gray circle represents the 0\farcs77 aperture at the protostellar position. \label{h2_single_temp}}
\end{figure*}

\begin{figure}[htbp!]
    \centering
    \includegraphics[width=0.5\textwidth]{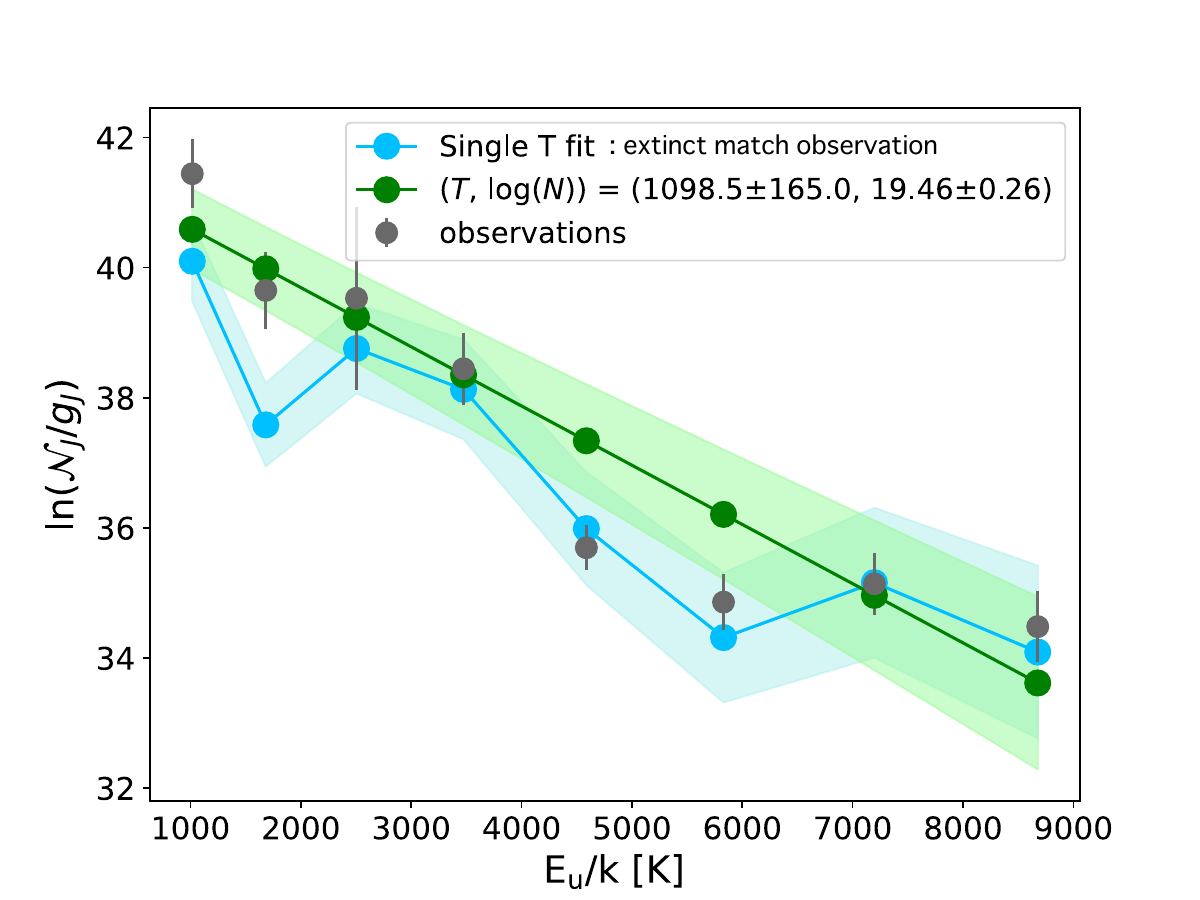}
    \caption{Rotation diagram at the continuum peak position shown by the ``x'' mark in Figure \ref{h2_single_temp}. The gray dots show the upper state column densities derived from the observed line fluxes, which suffer from extinction.  The blue connected dots show the best-fitting model single temperature model after taking into account the derived extinction, and the green connect dots show the same model with the extinction removed.  The values are extracted from a single pixel of 0\farcs13. \label{h2_rotation_diagram}}
\end{figure}

\subsection{Extinction Correction for H$_2$}
\label{sec:extinction}
The emission from an outflow is extincted by the ice and dust within the protostellar envelope as well as any foreground clouds.  Thus, to constrain the excitation of H$_2$ lines, the extinction needs to be corrected.  If we assume all H$_2$ emission arises from the outflow inside the envelope,
the effect of extinction can be described as
\begin{equation}
  I_{\text{obs}, \lambda_J, J_{\rm u}\rightarrow J_{\rm l}} = I_{\lambda_J, J_{\rm u}\rightarrow J_{\rm l}} e^{-(\tau_\text{ice} + \tau_\text{dust})},
  \label{eq:extinction}
\end{equation}
where $I_{\text{obs}, \lambda_J, J_{\rm u}\rightarrow J_{\rm l}}$ is the observed integrated intensity of an H$_2$ line and $I_{\lambda_J, J_{\rm u}\rightarrow J_{\rm l}}$ is the intrinsic integrated line intensity without extinction.  The $\lambda_J$ indicates the wavelength of the H$_2$ line.  The optical depths due to ice and dust are written as $\tau_\text{ice}$ and $\tau_\text{dust}$, respectively.  

We constrain these two sources of optical depth separately.  The ice absorption can be measured from the JWST MIRI spectrum (5--28 \micron; \citealt{2022ApJ...941L..13Y}) after subtracting the silicate absorption, derived by $\tau_{\rm ice} = -\text{log}(F_\text{obs}/F_\text{baseline})$.  We use a slightly updated version of the spectrum toward the protostar presented by \citet{2022ApJ...941L..13Y} as the template.  Then, shown in Figure\,\ref{fig:spec_silicate}, we fit a fourth-order polynomial baseline along with silicate absorption spectra that include both pyroxene (Mg$_{0.7}$Fe$_{0.3}$SiO$_3$) and olivine (MgFeSiO$_4$).  The spectra of pyroxene and olivine are synthesized using \texttt{optool} \citep{2021ascl.soft04010D} with the data from \citet{1995A&A...300..503D}.  To simplify the analysis of the region, we further assume a fixed ice optical depth across the FoV of the MIRI MRS. This hypothesis is appropriate as long as the ice absorption is dominated by ice in the outer envelope.  Future testing and verification of this assumption for the ice distribution is needed; however, it is beyond the scope of this study.

The dust optical depth is obtained in the following way.
The infinitesimal increase in the dust optical depth along a line of sight, $d\tau_{\text{dust}, \lambda}$, is written as
\begin{equation}\label{eq:tau_dust001}
d\tau_{\text{dust}, \lambda} = \kappa_\lambda \rho\ dl,
\end{equation}
where $\kappa_\lambda$ is the extinction per unit mass (absorption + scattering) cross section per gram of dust (cm g$^{-1}$), $\rho$ the mass density (g cm$^{-1}$), and $dl$ the infinitesimal path length.
Thus, the optical depth to a source at distance $d$ is given as 
\begin{equation}\label{eq:tau_dust002}
\tau_{\text{dust},\lambda} = \kappa_\lambda \int_{0}^{d}\rho\ dl.
\end{equation}
As $\int_{0}^{d}\rho\ dl$ equals the dust mass column density $\Sigma_\text{dust}$ (g cm$^{-2}$), the dust optical depth is obtained as
\begin{equation}\label{eq:tau_dust1}
  \tau_{\text{dust}, \lambda} = \Sigma_\text{dust} \kappa_\lambda. 
\end{equation}
Thus, the extinction-corrected integrated line intensity (Equation\,\ref{eq:extinction}) can be obtained if the dust mass column density ($\Sigma_\text{dust}$) is known.  
The extinction value in the V band \av\ is defined by
\begin{equation}\label{eq:tau_dust003}
{\rm A}_{\rm V} \equiv - 2.5 \log\ \frac{f_{\rm V}}{f_{\rm V,0}}= - 2.5 \log\ \frac{I_{\rm V}}{I_{\rm V,0}}.
\end{equation}
Using the relation,
\begin{equation}
I_{\rm V} = I_{\rm V,0}\ {\rm exp}\ (-\tau_{\rm V}), 
\end{equation}
${\rm A}_{\rm V}$ can be written as 
\begin{equation}\label{eq:tau_dust2}
   {\rm A}_{\rm V} = 1.086\tau_{\rm V}, 
\end{equation}
where $\tau_{\rm V}$ represents the dust optical depth in the V band ($\lambda =$ 0.55 \micron).

\subsection{Dust Models}\label{sec:dust model}
\begin{figure}
    \centering
    \includegraphics[width=0.47\textwidth]{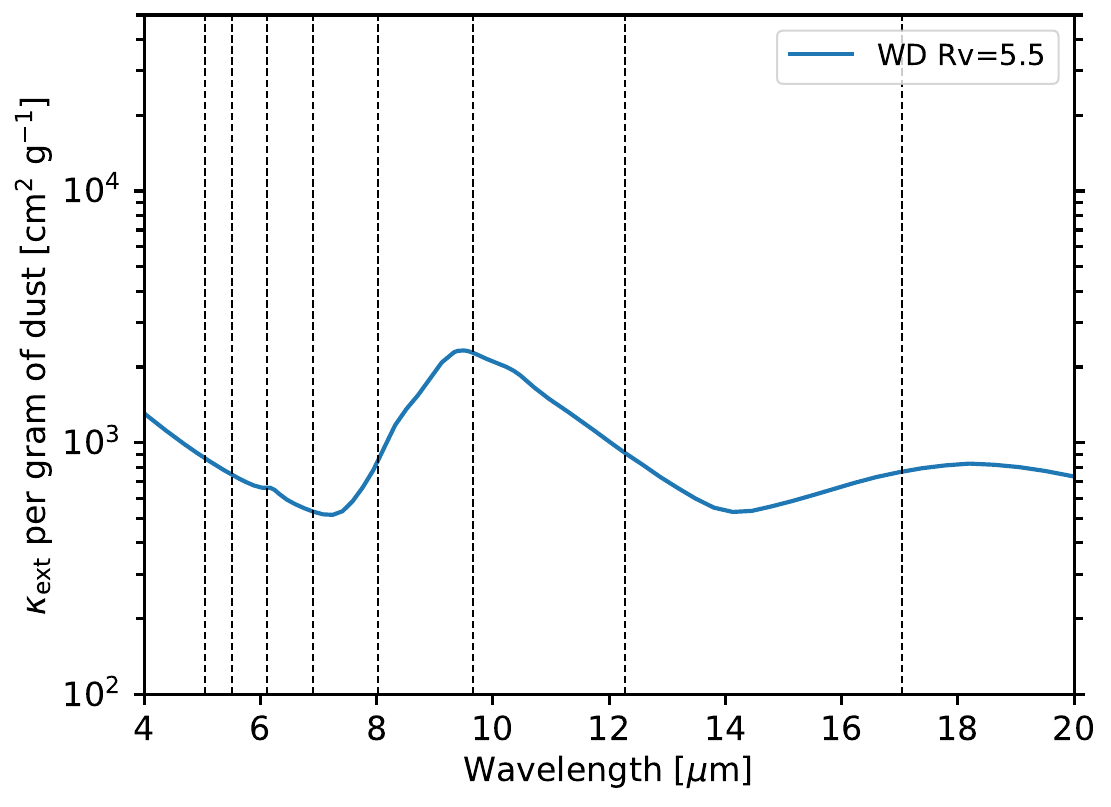}
    \caption{Extinction curve of $R_\text{V}=5.5$ from \citet{2001ApJ...548..296W}.  The dashed lines indicate the wavelengths of the H$_2$ lines.}
    \label{fig:wd_dust}
\end{figure}

In the main text of this paper, we mainly present the single-temperature fitting tested with the dust model from \cite{2001ApJ...548..296W}.
Their dust model employs $R_\text{V} = 5.5$ (hereafter WD5.5), as shown in Figure\,\ref{fig:wd_dust} , where $R_\text{V}$ represents $A_V/(A_B - A_V)$, which is the ratio of visual extinction to reddening \citep{2001ApJ...548..296W}.
Given that only a few H$_2$ lines are covered by MIRI MRS,single-temperature fitting is the simplest way to understand the distribution of physical parameters.  We choose a dust model without ice mantles for the main discussion, because we consider separately the ice absorption using the observed ice optical depth spectrum.

In Appendices \ref{ap:double} and \ref{ap:KP5}, we present double-temperature fitting using both the WD5.5 dust model as well as the KP5 dust model \citep{2024RNAAS...8...68P}.  As the double-temperature fitting has been adopted in several JWST observations of H$_2$ lines \citep{2024A&A...685C...5G, 2024ApJ...962L..16N, 2024A&A...687A..36T}, the results of our double-temperature fitting can be used for comparison to other sources.
In Section \ref{sec:comparison_temp}, we use the results of double-temperature fitting in \iras\ (Appendices \ref{ap:double} and \ref{ap:KP5}) to compare the source with another low-mass protostellar source IRAS 16253--2429 \citep{2024ApJ...962L..16N}. The double-temperature KP5 fitting aims to provide a reference for other studies where the KP5 dust model has been used.

Additionally, Appendix\,\ref{ap:JBB} presents the results of single-temperature fitting with a bare dust grain model developed based on the astro-silicate dust \citep{2003ApJ...598.1017D} used for the extinction estimate in \cite{2024ApJ...974...97S} (See also Section \ref{sec:single-icefree}).

\subsection{LTE Analysis of the H$_2$ Lines}
\label{sec:lte-H2}
\par Given $\tau_\text{ice}$ and $\tau_\text{dust}$, we can derive the intrinsic intensities of H$_2$ lines, $I_{\lambda_J, J_{\rm u}\rightarrow J_{\rm l}}$ (Equation\,\ref{eq:extinction}).  To calculate the column density in the upper state ($N_{J_{\rm u}}$), we assume local thermodynamic equilibrium (LTE) and relate the column density and the observed intensity as 
\begin{equation}
  N_{J_{\rm u}}=\frac{4\pi \lambda_{\rm u}}{hc}\frac{I_{J_{\rm u}\rightarrow J_{\rm l}}}{A_{J_{\rm u}\rightarrow J_{\rm l}}},
\end{equation}
where $A_{J_{\rm u}\rightarrow J_{\rm l}}$ is the Einstein A coefficient from $J_\text{u}$ to $J_\text{l}$ \citep{1998ApJS..115..293W}, $\lambda_\text{u}$ is the line wavelength \citep{1984ApJ...282L..85J}, $h$ is the Planck constant, and $c$ is the speed of light.
Assuming equilibrium conditions, the degeneracy $g=g_{s}(2J+1)$, where $g_{s}$ equals to 3 and 1 for ortho and para H$_2$, respectively, the relation between the column density in the ground state, $N$(H$_2$), and the excitation temperature (rotation temperature), $T$(H$_2$), is obtained by solving
\begin{equation}
  {\rm ln}\frac{N_{J_{\rm u}}}{g_{J_{\rm u}}} = -\frac{1}{T(\rm H_2) }\frac{E_{\rm u}}{k_{\rm B}}+ {\rm ln}\frac{N(\rm H_2)}{Q(T(\rm H_2))}.
\end{equation} 
where $E_{\rm u}$ and $k_{\rm B}$ are the upper-state energy and the Boltzmann constant, respectively.
$Q$($T$(H$_2$)) represents the partition function taken from \citet{1996AJ....111.2403H}, which is given as 
\begin{equation}
    Q(T({\rm H}_2)) =\frac{0.0247T({\rm H}_2)}{1-{\rm exp}(-\text{6000 K}/T({\rm H}_2))}. %
\end{equation}
To obtain $T$(H$_2$), $N$(H$_2$), and $\Sigma_{\rm dust}$ in the fitting, the optimization method used is the Trust Region Reflective algorithm along with the least squares statistic implemented in \texttt{scipy.optimize.least\_squares}.  



\subsection{Single-Temperature Fitting with the WD5.5 Dust Model}\label{sec:single-icefree}
\par 
In the single-temperature fitting with the WD5.5 dust model, we first determine $\Sigma_\text{dust}$ with the H$_2$ rotational diagram from the S(1) to S(5) lines, and then constrain $T$(H$_2$) and $N$(H$_2$) for the pixels where four or more lines are detected.
Thus, we exclude the higher excitation lines in determining $\Sigma_\text{dust}$, to avoid their scattering effects as much as possible.
The upper panels of Figure \ref{h2_single_temp} show the maps of $T$(H$_2$), $N$(H$_2$), and $\Sigma_{\rm dust}$ in the single-temperature fitting tested with the WD5.5 dust model, while the bottom panels are the uncertainty maps of each parameter.
We show the distributions of $T$(H$_2$) where the fitted value is three times greater than its uncertainty and those of $N$(H$_2$) and $\Sigma_{\rm dust}$  where the fitted values are greater than each fitting uncertainty.
The uncertainties of these three parameters originate from the uncertainties of the integrated H$_2$ intensities, which is each rms noise, and the 30 \% error of the mass extinction ($\kappa_{\rm ext}$) for each wavelength in the dust model. The rotation diagram at the protostar position for one pixel is shown in Figure \ref{h2_rotation_diagram}, as an example of the fitting.

The distribution of $T$(H$_2$) shows a concentrated hot spot centered at the north  near the protostar position, with the highest temperature reaching $\sim$ 1400 K (Figure \ref{h2_single_temp}(a)), although the uncertainties for these pixels are relatively large.
The region where the temperature is lower than 1100 K matches the distribution of the H$_2$ S(1) line (contours in Figure\,\ref{h2_single_temp}).
The higher temperature region is found outside the H$_2$ S(1) map.
Heating at this location could occur due to interaction between the envelope and the outflow.
In the southwestern outflow, the temperature gradually decreases from $\sim$ 1200 K to $\sim$900 K at $\sim2\arcsec$ from the protostar along the outflow direction (P.A. 235\degr).
The area with red dashed lines in Figure \ref{h2_single_temp}(a) represents the position of the first shell structure in the MIRI image (Figure \ref{miri}), and here we find that the temperature is 900--1000 K.
Throughout the outflow, the H$_2$ lines trace high-temperatures.

\par The CO temperature is measured by \cite{2024ApJ...974...97S} using the same JWST observation.
Their derived value is 1598$\pm$118 K within a 0\farcs77 aperture toward the protostar, where an \av\ of 15 mag ($\Sigma_{\rm dust}$=1.38$\times$10$^{-3}$ g$\cdot$cm$^{-2}$) was used for the extinction estimate in \cite{2024ApJ...974...97S} (Appendix \ref{ap:JBB}). We measure $T$(H$_2$) to be slightly lower, 1200$\pm$230 K.
The CO temperature is similar to the hotter component, 1389$\pm$83 K ($T_1$(H$_2$); Figure \ref{h2_double_temp_wd55}(a)), in the double-temperature fitting using the WD5.5 dust model (Appendix \ref{ap:double}), where the cooler component is significantly lower, 540$\pm$194 K ($T_2$(H$_2$); Figure \ref{h2_double_temp_wd55}(d)).

Using ALMA observed submillimeter \hhco\ lines, \cite{2020ApJ...900...40O} reported the kinetic temperature to be 54$\pm$2 K with an aperture of 0\farcs5 at the protostellar position.
In the same region, $T$(H$_2$) is derived to be 1147$\pm$198 K, much higher than the submillimeter \hhco\ temperature.
If the thermal temperature of H$_2$ is similar to its excitation temperature, the temperature difference between H$_2$ and \hhco\ implies that different layers of molecules are observed along the line of sight. H$_2$ mainly traces a hotter region, likely more inward of the envelope and the outflow cavity wall, whereas \hhco\ is emitting in the cooler part.
To confirm the \hhco\ temperatures in the outflow cavity wall, we estimate them at four positions, where the \hhco\ emission is enhanced, using recent ALMA Band 6 observation, as presented in Appendix \ref{ap:Tkin}. The temperatures found for \hhco\ are $<$260 K, which is roughly one order of magnitude lower than that of H$_2$.


Comparing to the outflow traced by the H$_2$ S(1) emission, the distribution of $N$(H$_2$) not only fills the same area but also extends slightly beyond the contours of the H$_2$ S(1) lines, with decreasing column density
(Figure \ref{h2_single_temp}(b)).  Comparing the outflow lobes, the southwestern lobe has higher column densities, approaching $\sim$10$^{20}$ cm$^{-2}$, while part of the northeastern lobe appears one order of magnitude lower ($\sim10^{19}$ cm$^{-2}$).
Within a 0\farcs77 aperture at the protostellar position, $N$(H$_2$) is measured to be (2.3$\pm$1.5)$\times$10$^{19}$ cm$^{-2}$. The $\Sigma_\text{dust}$ distibution (Figure \ref{h2_single_temp}(c)) is concentrated at the north of the protostar (2.1--2.5$\times$10$^{-3}$ [g cm$^{-2}$]) like $T$(H$_2$) (Figure \ref{h2_single_temp}(a)), and it has a relatively high mass column density of dust in the outflow cavity near the protostar as well ($\sim$2$\times$10$^{-3}$ [g cm$^{-2}$]). 

\subsection{Comparison with IRAS 16253--2429}\label{sec:comparison_temp}
Recently, the H$_2$ temperature has been reported for another low-mass protostellar source, IRAS 16253-2429, using H$_2$ S(1) to S(6) lines observed with JWST \citep{2024ApJ...962L..16N}, where the bolometric temperature is 42 K \citep{2023ApJS..266...32P}.
IRAS 16253-2429 has a very low bolometric luminosity (0.2 \sL) and is located in a relatively isolated region of the Ophiuchus molecular cloud ($d=$ 140 pc).
The mass of the central source was reported to be 0.12-0.17 \sm\ by \citet{2023ApJ...954..101A}.
We here compare the derived physical parameters in \iras\ with those in IRAS 16253-2429.

\cite{2024ApJ...962L..16N} fitted two temperature components toward the protostar, where the higher and lower temperatures are 849$\pm$104 K and 661$\pm$74 K, respectively.
For \iras, the two temperature fit yields a significantly higher hot component, 1389$\pm$83 K, and a comparable cool component, 540$\pm$194 K, to their results.
Note that the hotter component becomes much higher 3304$\pm$872 K ($T_1$(H$_2$); Figure \ref{h2_double_temp_kp5}(a)) in the double-temperature fitting using the KP5 dust model (Appendix \ref{ap:double}), where the cooler component is similar, 543$\pm$30 K ($T_2$(H$_2$); Figure \ref{h2_double_temp_kp5}(d)).
The high temperature difference might be caused by their evolutionary stage.
The mass accretion rates of IRAS 16253-2429 and \iras\ are reported to be (0.9-1.3)$\times$10$^{-7}$ \sm\ yr$^{-1}$ \citep{2023ApJ...954..101A} and (0.2-7.0)$\times$10$^{-6}$ \sm\ yr$^{-1}$ \citep{2014ApJ...795..152O, 2016AA...587A.145B, 2018ApJ...864L..25O, 2021AA...648A..41V}, respectively.
This suggests that IRAS 15398-3359 is in a more active accretion phase, which could cause the high temperature near the protostar.
Alternatively, the two temperature fitting may be influenced by the H$_2$ lines that are used, as we include higher excitation lines of H$_2$, S(7) and S(8), in our fitting.

Using the dust model from \citet{2024RNAAS...8...68P}, the optical depth (\av) is 26.5$\pm$0.5 mag at the protostar position in IRAS 16253-2429 \citep{2024ApJ...962L..16N}.  This result is about half that found for \iras\ (67$\pm$5 mag from a 0\farcs77 aperture).
This is consistent with expectations from circumstellar disk evolution. IRAS 16253-2429 has a developed disk \citep{2023ApJ...954..101A} in contrast to the \iras\ case \citep{2023ApJ...958...60T}. Thus, the inner envelope of \iras\ is less evolved and depleted, leading to denser conditions, and hence higher \av. 
We further note that, the envelope mass of IRAS 16253-2429 at the $\sim$ 1000 au scale is estimated with millimeter observations to be 0.2-1 \sm \citep[][]{2006A&A...447..609S, 2008ApJ...684.1240E, 2011ApJ...740...45T}, which is comparable or slightly lower than that of \iras\ at a similar or somewhat smaller scale  
\citep[0.5-1.2 \sm;][]{2012AA...542A...8K, 2013ApJ...779L..22J}.
Moreover, for both sources, \av\ decreases with increasing distance from the protostar.
The \av\ values were reported along the outflow of IRAS 16253-2429, linearly decreasing with distance from the protostar.
The blue-shifted and red-shifted outflows of IRAS 16253-2429 have 12.3$\pm$1.1 and 24.2 $\pm$ 0.7 at the edge of the FoV, respectively.
As a direct proxy for \av\, the $\Sigma_{\rm dust}$ map of \iras\, shown in Figure \ref{h2_single_temp}(c), also decreases as distance linearly in the southwestern outflow.

\section{Outflow and Jet Directions}
\label{sec:precession}
\par 
Variations in the position angle (P.A.) of the outflow and jet trace the flow over various timescales.  For \iras, \citet{2021AA...648A..41V} proposed an outflow precession scenario using sub-mm $^{12}$CO line emission.  They identified four ejections separated by 50--80 yrs in the southwestern outflow and reported their P.A. to be 214\degr, 227\degr, 236\degr, and 246\degr.  The P.A. slightly changes clockwise (smaller P.A.) toward the most recent ejection.  According to their study, the ejection with P.A.=214\degr\ has the fastest velocity ($\sim$10 \kms/sin $i_{\rm v}$) and shortest length (360 au/cos $i_{\rm v}$) among the four ejections  The $i_{\rm v}$ here represents the inclination angle of the ejection (24\degr; 0\degr\ for edge-on) employed in their analysis.

\begin{figure}[htbp!]
    \centering
    \includegraphics[width=0.47\textwidth]{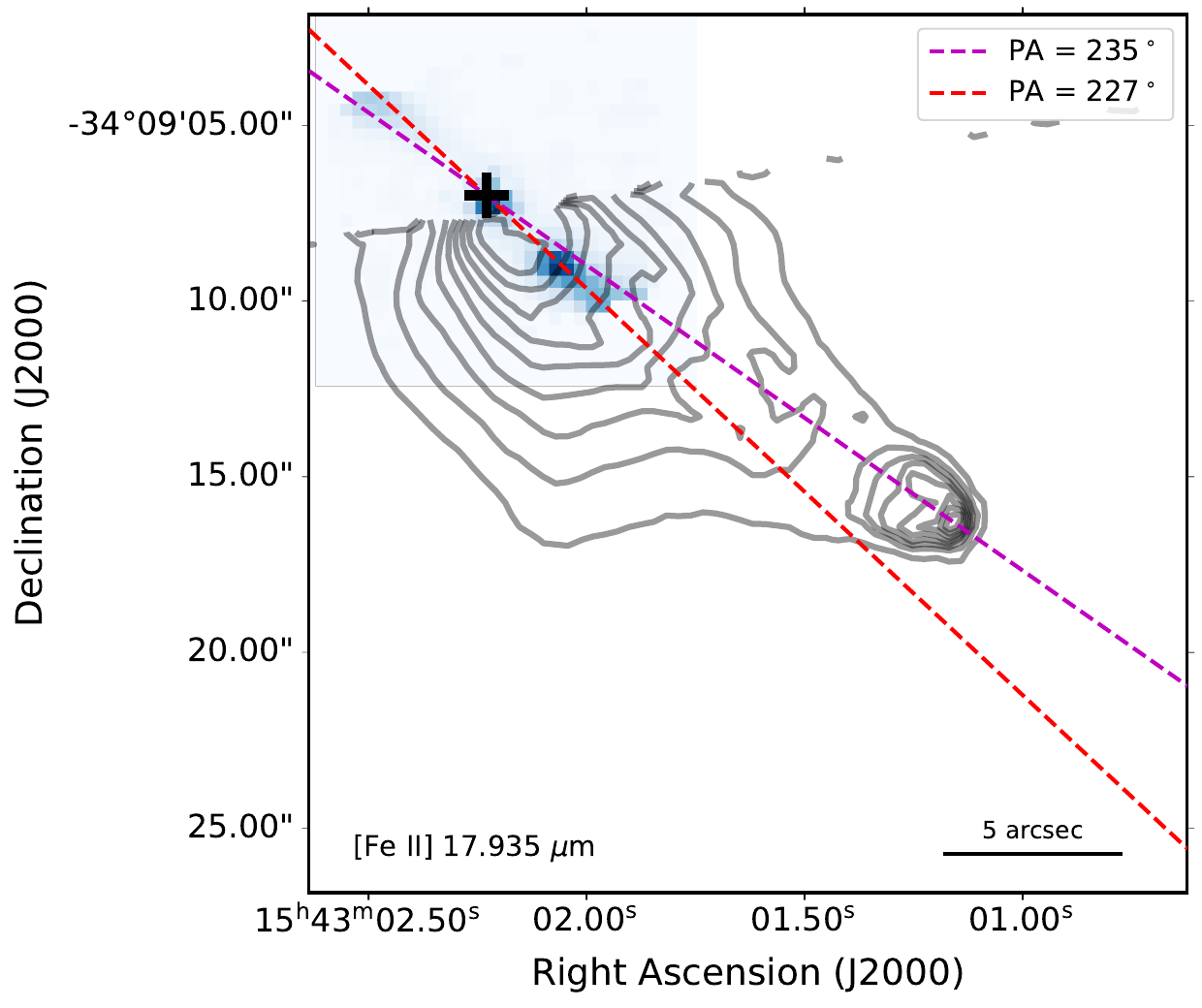}
    \caption{MIRI F560W image in contours overlaid with the image of the \feii\ 25.988 \micron\ line. The P.A. of the jet traced by \feii\ is 227.0\degr\ (red dashed lines), while that of the outflow in the MIRI image is 234.9\degr (purple dashed lines). The ``+'' marks the protostar position ($\alpha_{2000}$, $\delta_{2000}$) = 15$^{\rm h}$43$^{\rm m}$02$^{\rm s}$.232, $-$34\arcdeg09\arcmin06\farcs971. \label{PA_FE}}
\end{figure}

Figure \ref{PA_FE} shows the MIRI F560W continuum image  (contours) and the jet traced by the \feii\ 25.988 \micron\ line (color) to visualize the difference between their position angles.
The outflow in the MIRI image shows a P.A. of $234.9\degr\pm0.1\degr$, determined by the axis from the continuum peak to the center of the outermost shell (($\alpha_{2000}$, $\delta_{2000}$) = 15$^{\rm h}$43$^{\rm m}$01$^{\rm s}$.173$\pm$0.001, $-$34\arcdeg09\arcmin16\farcs205$\pm$0.001.), as shown in the purple dashed lines.  Note that the uncertainties of the P.A. are propagated from the uncertainties of the two positions. This P.A. is slightly smaller than the second oldest ejection (P.A.=236\degr) identified by \citet{2021AA...648A..41V} but greater than the outflow P.A. previously reported using ALMA observations, 230\degr\ \citep{2014ApJ...795..152O} and 220\degr\ \citep{2018ApJ...864L..25O}.
The discrepancy of P.A. could be due to the non-detection of compact sub-mm molecular emission at the end of the outflow for an accurate measurement of P.A.  
Note that the dynamical timescale was reported to be 170 yrs for the outermost shell in the MIRI image using previous Spitzer data to deduce the proper motion \citep{2022ApJ...941L..13Y}, which is reasonably consistent with the 221 yr for the second oldest ejection seen by \citet{2021AA...648A..41V}.
We also detect extended scattered light in the MIRI image, which could imply a wide opening angle for the outflow, potentially due to either multiple ejections with changing direction or a continually precessing outflow.


Within the outflow, the jet has a P.A. of 227.0\degr\ $\pm$1.1\degr, determined by the axis from the continuum peak to the peak of \feii\ in the southwest (($\alpha_{2000}$, $\delta_{2000}$) = 15$^{\rm h}$43$^{\rm m}$02$^{\rm s}$.042$\pm$0.051, $-$34\arcdeg09\arcmin09\farcs162$\pm$0.040.).
This direction is offset from that seen in the MIRI F560W image, but has the same P.A. as the second youngest ejection reported by \citet{2021AA...648A..41V}.  The jet is measured at a much smaller scale compared to the outflow in the MIRI image, and therefore likely traces more recently ejected material.  
We estimate the timescale of this jet knot using the line-of-sight \feii\ velocity of 89 km$\cdot$ s$^{-1}$ derived from the line fitting at the southwest peak of \feii, which is much faster than the $^{12}$CO outflow observed with ALMA ($<$10 \kms).
The jet velocity and the distance from the protostar are corrected to be 89 km$\cdot$ s$^{-1}$/sin $i$ and 500 au/cos $i$, respectively, by the inclination angle ($i$).
The timescale is then estimated to be 10 yrs, in the case that we employ the inclination angle ($i$) of 20\degr (0\degr\ for edge-on). This inclination is used by the previous studies \citep{2014ApJ...795..152O, 2018ApJ...864L..25O}, and similar to that employed by \cite{2021AA...648A..41V}.
Even without corrections for the inclination, the timescale (26 yrs) is shorter than the primary outflow dynamical timescale mentioned in the previous paragragh \cite[170 yrs;][]{2022ApJ...941L..13Y}.
As well, it is shorter than the dynamical timescales derived from the gas emission reported previously (62--268 yrs; \citealt{2021AA...648A..41V}, $\sim$1000 yrs; \citealt{2014ApJ...795..152O}).
Thus, the [Fe II] knot seems to be the most recently ejected material, in relation with the other observed emission from this source.

One of the outflow ejections identified by \cite{2021AA...648A..41V} has a smaller P.A. of 214\degr.
If the \feii\ emission is more recent than the most recent CO ejection, then the outflow may have moved back toward a P.A. of 227\degr\ ([Fe II] knot), after having moved to P.A of 214\degr\ from a P.A. of 227\degr\ (second most recent CO ejection).
The reversal in the P.A. implies a directional change.
More observations are required to confirm this potential change of the jet precession.
The motion of the source itself with respect to its environment can contribute to the outflow and jet structures as well, as has been found in another low-mass protostellar source B335 \citep{2024AJ....167..102H}.
In that case the jet may appear to bend as different forces act on the jet versus the central source.

\section{Summary}\label{sec:summary}
We have studies the outflow/jet structure of the Class 0 low-mass protostellar source \iras\ using observations from JWST and ALMA. The main results are summarized below.\\

\begin{enumerate}
    \item We newly report the RGB MIRI image, the H$_2$ (S(1)-S(8) maps corrected for extinction, and the \feii\ (5.34 \micron) and \niii\ (6.64 \micron) maps. 

    \item  To understand the outflow morphology, we compare the MIRI F560W image showing four shell structures in the southwestern outflow at a few 1000 au scale to the ALMA molecular-line maps. The first and second shells are also seen in the \hhco, CS, CCH, and \meta\ lines. The third one is barely seen in the \hhco\ and CS lines. \ccchh\ traces the first shell.  None of the molecular lines observed with ALMA trace the outermost shell seen in the MIRI image, where the MIRI RGB image observed with JWST reveals a blue tip. This suggests that the outermost shell lacks a component cool enough for ALMA to detect, even for shock tracers (\meta, SO, and SiO).

    \item Pure rotational lines of H$_2$ (S(1)-S(8)) and ionized lines (\feii, \neii, \si, and \niii) are observed within a few hundred au scale around the protostar. The H$_2$ lines show the outflow structure, whereas the \feii\ (25 \micron\ and 17 \micron) and \neii\ show more collimated structures coming from a jet. We find that the \si\ map is comparable to that of H$_2$ S(1), and that \feii\ (5 \micron) and \niii\ show compact emission next to the protostar. The H$_2$ S(1) line and the other ion lines are also enhanced at this position, suggesting a jet launching point.
    
    \item We compare the H$_2$ S(1) map to the $^{12}$CO, \hhco, CS, CCH, and \ccchh\ lines observed with ALMA. The outline observed with H$_2$ is consistent with the ALMA images. The molecular lines from ALMA mainly trace the cavity and the shell, while the H$_2$ S(1) line shows a strong peak inside the cavity. 
    
    \item By using the 8 lines of H$_2$, we derive the excitation temperature, column density of H$_2$, and the mass column density of dust (visual extinction) for each pixel. In the single-temperature fitting with the WD5.5 dust model \citep{2001ApJ...548..296W}, these values from a 0\farcs77 aperture toward the protostar are derived to be 1200$\pm$230 K, (2.3$\pm$1.5)$\times$10$^{19}$ cm$^{-2}$, and 2.0$\pm$0.1 g$\cdot$cm$^{-2}$ (67$\pm$5 mag), respectively. Toward the southwest, the temperature goes down to $\sim$900 K at $\sim$2$''$ from the protostar along the outflow direction (P.A. 235\degr). Within 0\farcs5 of the protostar, the H$_2$ temperature (1147$\pm$198 K) is much higher than the \hhco\ temperature measured with ALMA previously \citep[54$\pm$2 K;][]{2020ApJ...900...40O}. Thus, we find that a hot region which cannot be detected in molecular lines with radio observations is well traced by H2 with JWST.

    \item We measure the outflow direction from the protostar to the outermost shell in the MIRI image of F560W and the jet direction from the protostar to the peak further from the protostar in \feii\ (25\micron) to be 234.9\degr\ $\pm$0.1\degr and 227.0\degr\ $\pm$1.1\degr, respectively.
    The jet is likely ejected very recently, based on the short dynamical timescale (10 yrs).
    While this difference might be due to the jet precession, 
    future observations are necessary to understand the mechanism.
\end{enumerate}

\begin{acknowledgments}
The authors are grateful to Ewine F. van Dishoeck for useful discussion.
This work is based on observations made with the NASA/ESA/CSA James Webb Space Telescope. The data
were obtained from the Mikulski Archive for Space Telescopes at the Space Telescope Science Institute, which is operated by the Association of Universities for Research in Astronomy, Inc., under NASA contract NAS 5-03127 for JWST.
These observations are associated with JWST GO Cycle 1
program ID 2151. 
The data presented in this paper were obtained from the Mikulski Archive for Space Telescopes (MAST) at the Space Telescope Science Institute. The specific observations analyzed can be accessed via \dataset[doi: 10.17909/wv1n-rf97]{https://doi.org/10.17909/wv1n-rf97} and \dataset[doi: 10.17909/qv17-1b93]{https://doi.org/10.17909/qv17-1b93}.
This paper makes use of the following ALMA data set:
ADS/JAO.ALMA\# 2013.1.00879.S. (PI: Hsi-Wei Yen), 2018.1.01205.L. (PI: Satoshi Yamamoto), 2019.1.01359.S (PI: Yuki Okoda). ALMA is a partnership of the ESO (representing its member states), the NSF (USA) and NINS (Japan), together with the NRC (Canada) and the NSC and ASIAA (Taiwan), in cooperation with the Republic of Chile.
The Joint ALMA Observatory is operated by the ESO, the AUI/NRAO, and the NAOJ. The authors thank to the ALMA staff for their excellent support.
This project is supported by a Grant-in-Aid from Japan Society for the Promotion of Science (KAKENHI: No. 20H05845, 20H05844, 22K20389, 22K20390.) and a pioneering project in RIKEN (Evolution
of Matter in the Universe).
Y. Okoda thanks RIKEN Special Postdoctoral Researcher Program (Fellowships) for financial support.
D.J.\ is supported by NRC Canada and by an NSERC Discovery Grant.
\end{acknowledgments}

%

\vspace{5mm}
\facilities{JWST, ALMA}



\bibliography{main}
\bibliographystyle{aasjournal}

\appendix
\setcounter{figure}{0}
\setcounter{table}{0}
\section{Observed H$_2$ maps}\label{ap:H2maps}
\begin{figure*}[htbp!]
    \centering
    \includegraphics[width=\textwidth]{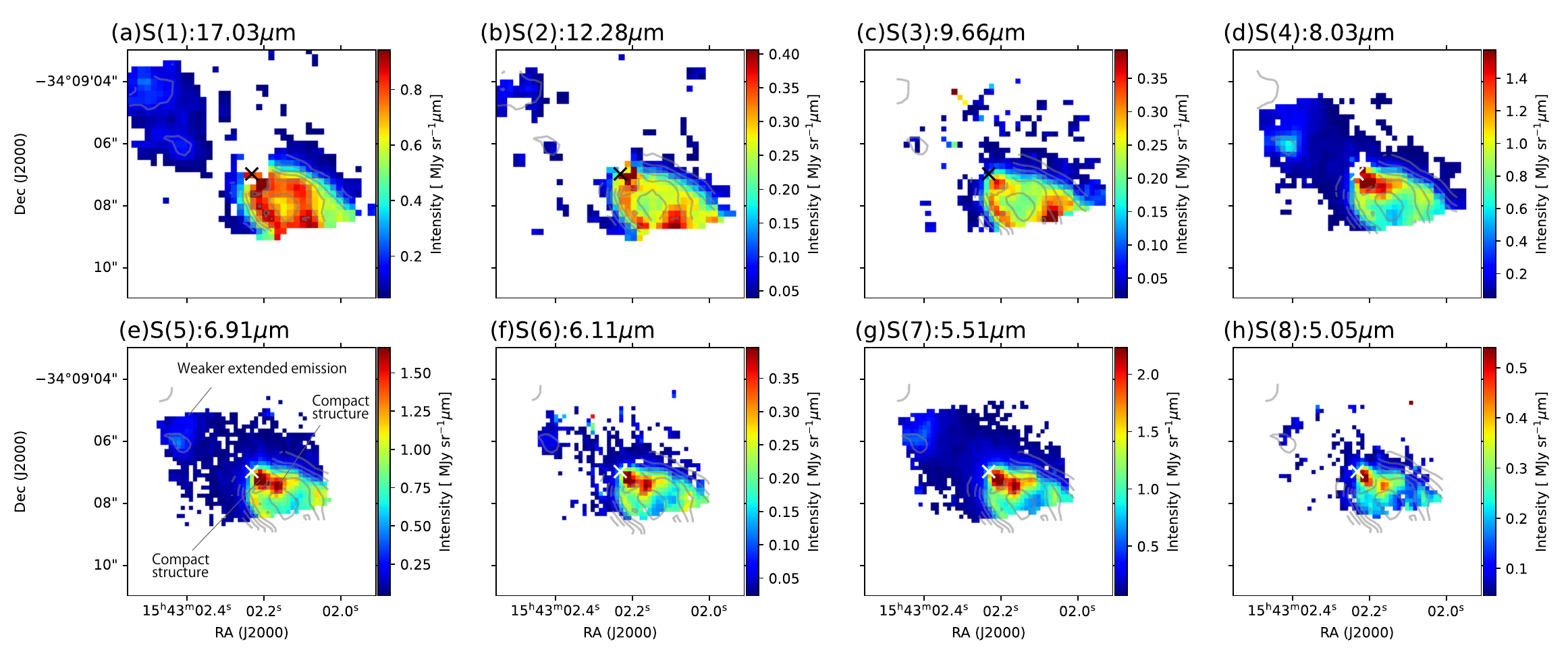}
    \caption{Integrated intensity maps of H$_2$ lines. Each peak intensity and FWHM are obtained from the Gaussian fitting after continuum subtraction. The pixels have an intensity higher than 3 $\sigma$ noise level. The ``x'' marks the protostellar position, ($\alpha_{2000}$, $\delta_{2000}$) = 15$^{\rm h}$43$^{\rm m}$02$^{\rm s}$.232, $-$34\arcdeg09\arcmin06\farcs971. Gray contours show the map of the H$_2$ S(1) line. Contour levels are every 3$\sigma$ from 3$\sigma$, where $\sigma$ is 0.05 MJy sr$^{-1}$ \micron. \label{fig:h2wex}}
\end{figure*}

Figure \ref{fig:h2wex} shows integrated intensity maps of H$_2$ lines from S(1) to S(8) without extinction correction or re-gridding.  The maps of S(5) and S(7) were reported previously as well \citep{2022ApJ...941L..13Y}. We obtained these images by fitting a Gaussian profile with a linear continuum.
The FoV of each line is determined by the integral field unit (IFU) used for the observations.  The FoV is larger at longer wavelengths.
 
\counterwithin{figure}{section}
\counterwithin{table}{section}

\newpage
\section{Double-temperature fitting of H$_2$ (WD5.5 Dust Model)}\label{ap:double}

We here present a double-temperature fitting with the WD5.5 dust model \citep{2001ApJ...548..296W}.
In this fitting, we use our ice optical depth spectrum (Figure \ref{fig:spec_silicate}) to obtain the extinction contribution of ice. 
We calculate two temperatures ($T_1$(H$_2$) and $T_2$(H$_2$)), two column densities ($N_1$(H$_2$) and $N_2$(H$_2$)), and the mass column density of dust ($\Sigma_{\rm dust}$)) for each pixel where six or more lines are detected.
$\Sigma_{\rm dust}$ is obtained by the fitting in the excitation modeling of the H$_2$ lines (Section\,\ref{sec:lte-H2}), as well as the temperature and the column density of H$_2$ simultaneously.

Figure \ref{h2_double_temp} shows the maps of the five parameters and each uncertainty.
For the temperatures, we show the maps where the fitted values are three times greater than the fitting uncertainties.
For the others, the fitted values greater than the fitting uncertainties are shown.
The southwestern outflow has double components where the S(1) line map of H$_2$ covers.
Both of $T_1$(H$_2$) and $T_2$(H$_2$) tend to decrease from 1400 K to 1100 K and from 600 K to 550 K toward the southwest, respectively. 
$T_1$(H$_2$) has a higher temperature, roughly 1500 K, outside the area covered by the S(1) line.
At the continuum peak, $T_1$(H$_2$) and $T_2$(H$_2$) are derived to be 1384$\pm$217 K and 588$\pm$90 K, respectively.
The $T_1$(H$_2$) value at the protostar is similar to the $T$(H$_2$) derived using a single-temperature fitting (1098$\pm$165 K; Figure \ref{h2_rotation_diagram}) within the uncertainty.
While the absolute values of $T_1$(H$_2$) are higher, the distribution of $T_1$(H$_2$) is also comparable to that of $T$(H$_2$) derived in a single-temperature fitting (Figure \ref{h2_single_temp}(a)).
For the column densities, we can see a boundary which is consistent with the map of the S(1) line. $N_1$(H$_2$) shows a higher column density up to $\sim$2.0$\times$10$^{19}$ cm$^{-2}$ in the southwestern outflow.
On the other hand, $N_2$(H$_2$) has a concentrated structure to the north of the protostar and has a extrema at the peak of the S(1) line which is 0\farcs5 from the protostar toward the southwest.
The values of $N_2$(H$_2$) are higher than those of $N_1$(H$_2$).
At the continuum peak, $N_1$(H$_2$) and $N_2$(H$_2$) are (8.9$\pm$5.6)$\times$10$^{18}$ cm$^{-2}$ and (1.3$\pm$0.4)$\times$10$^{20}$ cm$^{-2}$, respectively.
$\Sigma_{\rm dust}$ looks similar to that in the single-temperature fitting (Figure \ref{h2_single_temp}(c)), although the values in the double-temperature fitting are a little lower.

The double-temperature fitting suggests that the outflow might have two quite different components 
along the line of sight, although the parameters have large uncertainties at some pixels.
In either of single and double-temperature fitting, we find that a hot region is traced by H$_2$ lines.

\begin{figure*}[htbp!]
  \centering
  \includegraphics[width=\textwidth]{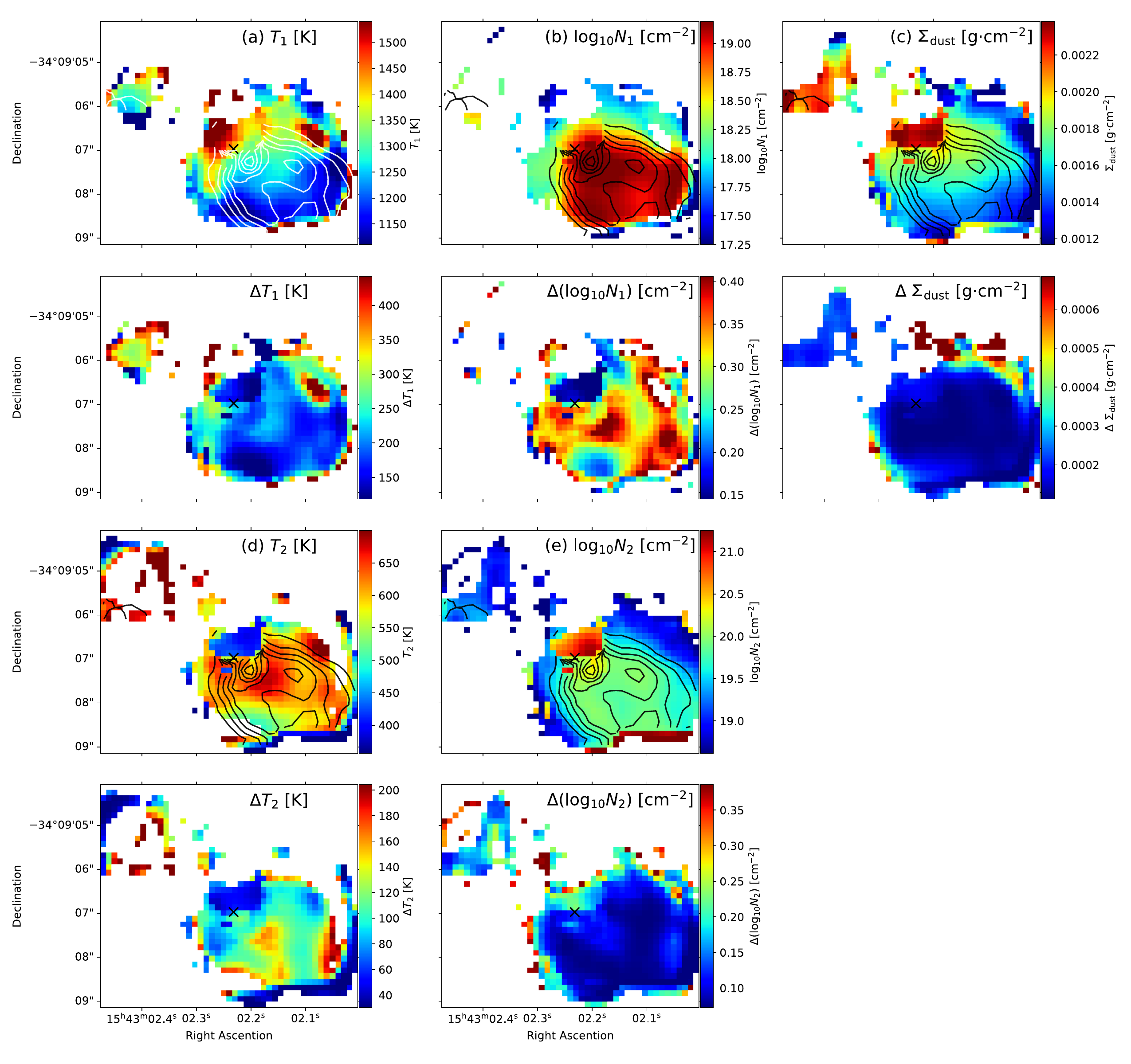}
  \caption{Results of double-temperature fitting using the WD5.5 dust model \citep{2001ApJ...548..296W}. (a, b) The temperature and column density maps of H$_2$ for the first (hotter) components, respectively. (c) Mass column density of dust.
  (d, e) The temperature and column density maps of H$_2$ for the second (cooler) components, respectively. The two temperature and two column density values are calculated at the pixels where the five or more lines are detected. 
  We here show the distributions where each value is higher than each uncertainty. Contours show the H$_2$ S(1) map shown in Figure \ref{h2_line}.
  Bottom images of each panel show the uncertainties. The ``x'' marks the protostellar position. \label{h2_double_temp_wd55}}
\end{figure*}

\newpage
\section{Double-temperature fitting of H$_2$ (Icy Dust Model)}\label{ap:KP5}
\begin{figure}[htbp!]
  \centering
  \includegraphics[width=0.6\textwidth]{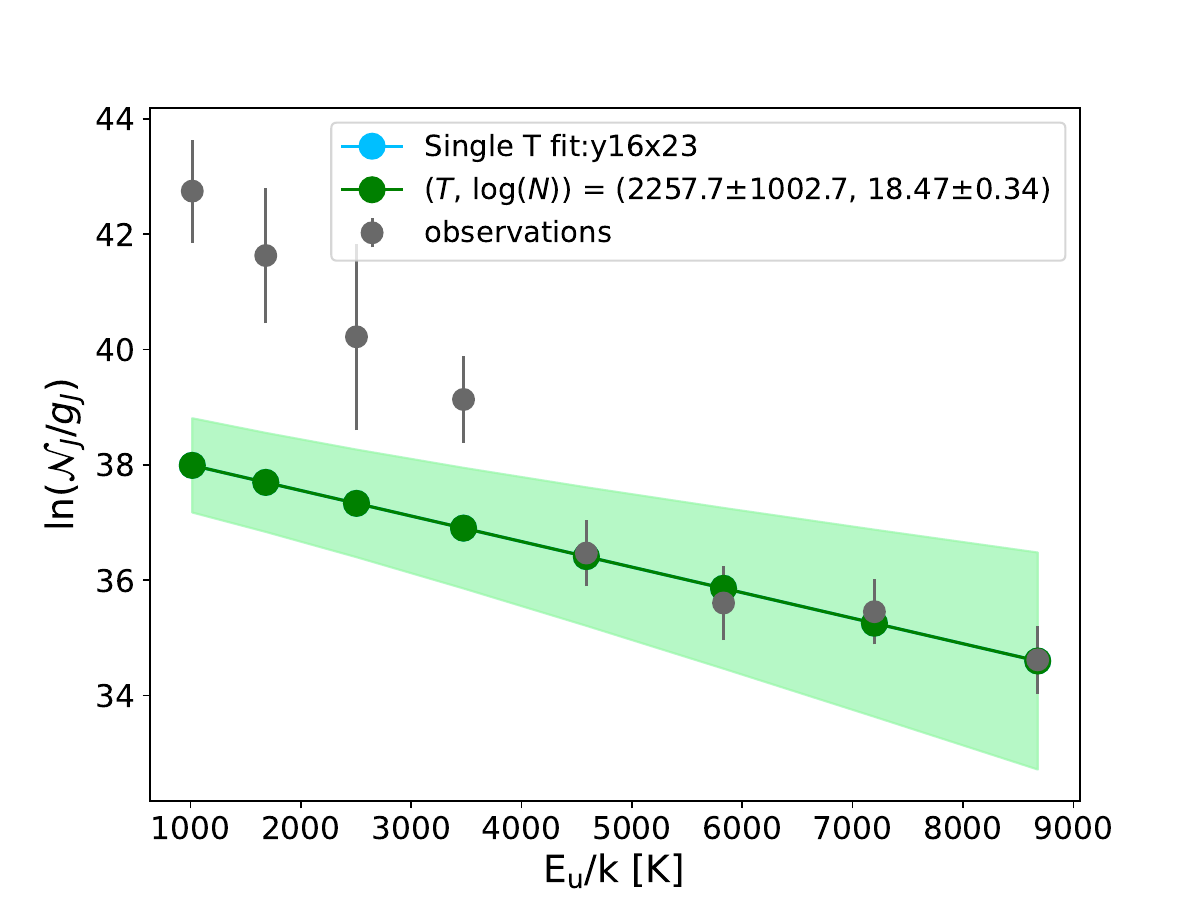}
  \caption{Rotational disgram of single-temperature fitting at the continuum peak when the icy dust model KP5 is employed. Same as Figure \ref{h2_rotation_diagram}. \label{kp5_rotation}}
\end{figure}

\begin{figure*}[htbp!]
  \centering
  \includegraphics[width=\textwidth]{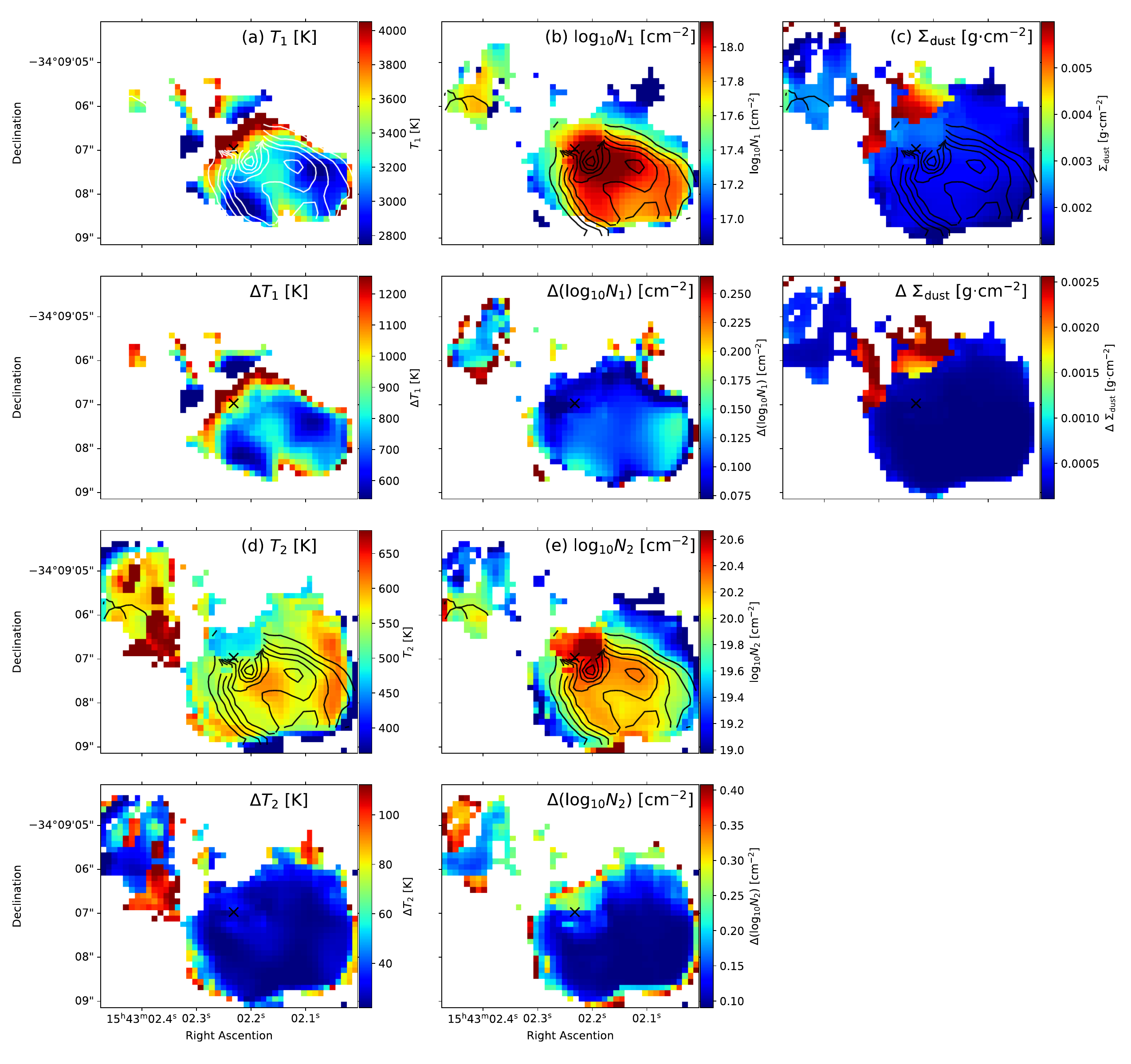}
  \caption{Results of double-temperature fitting when the icy dust model KP5 is employed. (a-e) Same as Figure \ref{h2_double_temp_wd55}. \label{h2_double_temp_kp5}}
\end{figure*}

\begin{figure*}[htbp!]
  \centering
  \includegraphics[width=\textwidth]{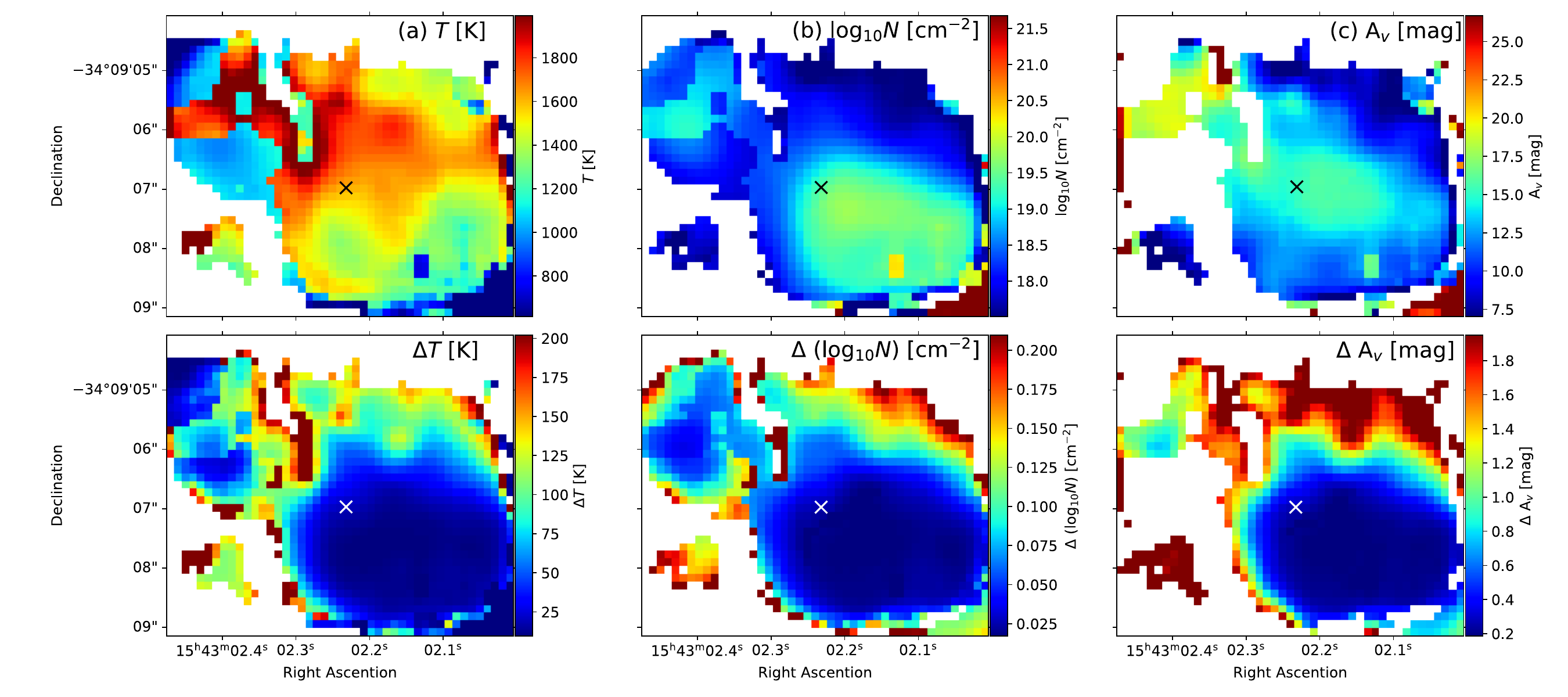}
  \caption{(a, b) The temperature and column density maps of H$_2$, respectively.(c) Visual extinction, \av.
  We show the distributions where the temperature value is greater than three times uncertainty. For the others, the distributions are shown where the value is greater than each uncertainty.
  Bottom images of each panel show the uncertainties. The ``x'' marks the protostellar position. \label{h2_double_temp}}
\end{figure*}

In order to compare with the results of double-temperature fitting using the WD5.5 dust model \citep{2001ApJ...548..296W}, we also test single and double-temperature fitting using the icy dust model KP5 \citep{2024RNAAS...8...68P}.
We first test a single-temperature fitting in the same way as single-temperature fitting with the WD5.5 dust model (Section \ref{sec:single-icefree}).
Figure \ref{kp5_rotation} shows the rotational diagram at the continuum peak, and we find that the first four lines are not fitted well with a single component.
Hence, we here present a double-temperature fitting with the model KP5.
The results are presented in Figure \ref{h2_double_temp_kp5}.

$T_1$(H$_2$) shows a much higher temperature with large uncertainty, compared to that with the WD5.5 dust model (Figure \ref{h2_double_temp_wd55}(a)).
In the $T_2$(H$_2$) map, the values are similar to those with the WD5.5 dust model (Figure \ref{h2_double_temp_wd55}(d)), which is roughly 500--600 K.
At the continuum peak, these values are derived to be 3767$\pm$978 K and 523$\pm$27 K, respectively.
As shown in Figures \ref{h2_double_temp_kp5}(b) and (e), both $N_1$(H$_2$) and $N_2$(H$_2$) are comparable to those with the WD5.5 dust model (Figures \ref{h2_double_temp_wd55}(b) and (e)).
The values at the continuum peak are (1.2$\pm$0.2)$\times$10$^{18}$ cm$^{-2}$ and (3.0$\pm$1.0)$\times$10$^{20}$ cm$^{-2}$, respectively.
$\Sigma_{\rm dust}$ also looks similar to that with the WD5.5 dust model, especially except for the northeastern outflow.

\newpage
\section{Single-temperature fitting of H$_2$ (Bare Dust Grain Model)}\label{ap:JBB}
In addition to the published dust models, we also explored synthetic dust models that are tailored for this source.  Here we presented the extinction and excitation fitting results using a dust model developed for radiative transfer modeling of this source.  While this particular dust model is no longer used for the modeling projects because of a more versatile model was made available, the extinction fitted using this dust model is used for the analysis of water and CO lines presented in \citet{2024ApJ...974...97S}.  Thus, we present the fitting results here to maintain consistency with related studies.

We used the dust model presented in \citet[][Fig. 4]{2024ApJ...974...97S} to fit for a single temperature component.
With a 0\farcs{77} aperture at the protostar, we derived a visual extinction of \av=15 mag ($\Sigma_{\rm dust}$=1.38$\times$10$^{-3}$ g$\cdot$cm$^{-2}$).
We obtain $T$(H$_2$), $N$(H$_2$), and $\Sigma_{\rm dust}$ simultaneously, by the fitting in the excitation modeling of the H$_2$ lines (Section\,\ref{sec:lte-H2}).
At the continuum peak,  $T$(H$_2$) and $N$(H$_2$) are derived to be 1606$\pm$20 K, (2.7$\pm$0.1)$\times$10$^{19}$ cm$^{-2}$, respectively.
Compared to the results with the WD5.5 dust model shown in Figure \ref{h2_single_temp} (see also Section \ref{sec:single-icefree}), $T$(H$_2$) and $N$(H$_2$) with a bare dust grain model are slightly higher and lower, respectively.
Distributions of $T$(H$_2$) and $N$(H$_2$) also have a little difference from Figures \ref{h2_single_temp}(a) and (b), 
although our conclusions do not change whichever model we employ.

\newpage

\section{Outflow Cavity Properties Derived from ALMA \hhco}\label{ap:Tkin}
\par 
We here study the southwestern outflow cavity to estimate the kinetic temperature of \hhco.
The three \hhco\ lines, 3$_{0,3}-$2$_{0,2}$, 3$_{2,1}-$2$_{2,0}$, and 3$_{2,2}-$2$_{2,1}$, were recently observed by ALMA (2019.1.01359.S) at a high resolution of $\sim$0$\farcs$1.
The FoV covers the area where the four shell structures are seen in the MIRI image. 
The line and observation parameters are summarized in Tables \ref{ALMA_observations_2019} and \ref{parameter_observation}, respectively.

\begin{table*}[htbp!]

  \caption{Line Parameters Observed with ALMA (2019.1.01359.S). \label{ALMA_observations_2019}}
  \centering
  \begin{tabular}{cccccc}
  \hline
  Line  & Transition & Frequency  &  Beam size  \\
        &            & (GHz)      &             \\
  \hline
  \hhco     & 3$_{0,3}-2_{0,2}$                   & 218.2221920           & 0\farcs13$\times$0\farcs10 (P.A. = 87.3\degr)   \\
                & 3$_{2,1}-2_{2,0}$                   & 218.7600660           & 0\farcs13$\times$0\farcs10 (P.A. = 87.6\degr)  \\
                & 3$_{2,2}-2_{2,1}$                   & 218.4756320           & 0\farcs13$\times$0\farcs10 (P.A. = 87.3\degr)     \\
  \hline
  \end{tabular}
  \begin{flushleft}
  \end{flushleft}
\end{table*}

\begin{table*}[htbp!]
  \caption{Observation Parameters for ALMA Program 2019.1.01359.S \label{parameter_observation}}
  \centering
  \begin{tabular}{lcccc}
  \hline \hline
  Execution block &1$^{a}$&2$^{b}$&3$^{c}$&4$^{d}$	\\
  \hline
  Observation date			                  & 2021 July 18  & 2021 July 20  & 2021 July 21  & 2021 July 26-27 \\
  Time on Source (minute)	                & 49.22         & 49.30         & 49.22         & 49.33           \\
  Number of antennas			                & 43            & 48            & 48            & 38              \\
  Observation frequency (GHz)             & \multicolumn{4}{c}{215.2 - 219.4}\\
  Maximum recoverble scale  (\arcsec)	    & 1.8           & 1.7           & 1.7           & 1.4						  \\
  Total bandwidth (GHz) 		              & \multicolumn{4}{c}{0.059} \\
  Spectral channel width (MHz)            & \multicolumn{4}{c}{0.122} \\
  Continuum bandwidth (GHz)	              & \multicolumn{4}{c}{1.88}  \\
  Baseline range (m) 		                  & 15-3638.2   & 15-3696.9	      & 15-3696.9     & 15-3321.0		    \\
  Bandpass calibrator			                & J1517$-$2422& J1427$-$4206    & J1924$-$2914  & J1427$-$4206	  \\
  Phase calibrator				                & \multicolumn{4}{c}{J1534$-$3526}	\\
  Flux calibrator				                  & J1517$-$2422& J1427$-$4206    & J1924$-$2914  & J1427$-$4206	  \\
  Pointing calibrator			                & J1517$-$2422& J1427$-$4206    & J1457$-$3539, J1924$-$2914 & J1427$-$4206 \\		
  RMS$^{e}$ (mJy beam$^{-1}$channel$^{-1}$)& \multicolumn{4}{c}{1.6} \\
  \hline
  \end{tabular}
    \begin{flushleft}
   \tablecomments{$^{a}${uid\_\_\_A002\_Xee1eb6\_X1912}
  $^{b}${uid\_\_\_A002\_Xee1eb6\_X12299}
  $^{c}${uid\_\_\_A002\_Xee1eb6\_X132c0}
  $^{d}${uid\_\_\_A002\_Xee7674\_X41f}
  $^{e}${Root mean square noise per one channel.}}
    \end{flushleft}
  \end{table*}

\subsection{Distribution and Spectrum}
\par
\begin{figure*}[htbp!]
  \centering
  \includegraphics[scale=0.22]{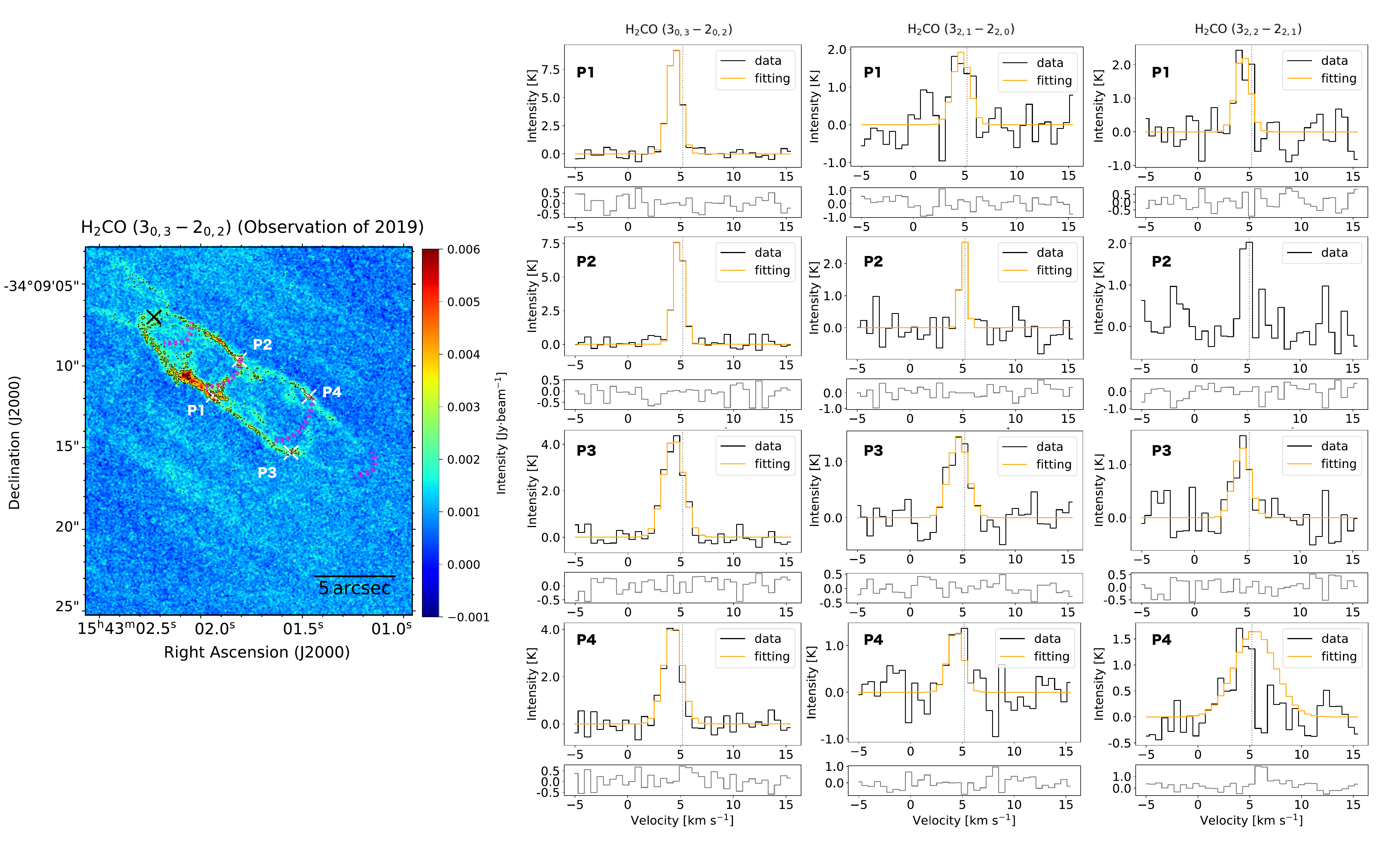}
  \caption{Left: Peak intensity (Moment 8) maps of \hhco\ (3$_{0,3}-$2$_{0,2}$) observed with ALMA observations, 2019.1.01359.S. Contours represent every 3$\sigma$ from 3$\sigma$, where $\sigma$ are 1\mjybeam. Black ``x'' mark shows the protostar position (($\alpha_{2000}$, $\delta_{2000}$)= 15$^{\rm h}$43$^{\rm m}$02$^{\rm s}$.232, $-$34\arcdeg09\arcmin06\farcs971), and white ``x'' marks show the positions of P1, P2, P3, and P4, which are summarized in Table \ref{hhco_temp}. Right: Spectra of the \hhco\ lines toward P1--P4.  Each panel shows the spectrum extracted from a 0\farcs{5} aperture (black) along with the fitted Gaussian profile (orange).  The residual after subtracting the fitted Gaussian profile is shown beneath each spectrum. Dotted vertical lines represent the systemic velocity of 5.2 \kms \citep[e.g.,][]{2017ApJ...834..178Y}.\label{spectra_hhco}}
\end{figure*}

 
To derive the temperature in the outflow cavity, we use the \hhco\ lines observed in ALMA program 2019.1.01359.S because this observation contains three lines of \hhco. 
Left panel of Figure \ref{spectra_hhco} shows the peak intensity (moment 8) map of the \hhco\ 3$_{0,3}-$2$_{0,2}$ line toward the southwestern outflow, observed in the 2019 program, where the protostar position is shown as the black ``x'' mark in the upper-left corner.
This observation has a much smaller maximum recoverable scale of 1\farcs0, compared to the FAUST data used for morphological comparison in Section\,\ref{sec:comparison_large} (Figure \ref{moment_alma}(b)), which leads to some emission being resolved-out.  The details for the FAUST observations are presented in Section \ref{sec:ALMAobs}.

The \hhco\ line (Figure \ref{spectra_hhco}) traces several emission features along the outflow cavity wall.
We identify the four positions (P1, P2, P3, and P4), corresponding near the edges of the second and third shell structures in the MIRI images (Figure \ref{miri}).  
Right panel of Figure \ref{spectra_hhco} shows the observed spectra of the three lines at these positions along with the fitted Gaussian profile.
We employ the systemic velocity of 5.2 \kms\ \citep[e.g.,][]{2017ApJ...834..178Y}.
The fitting results are summarized in Table \ref{line_para}.
Overall, the 3$_{0,3}-$2$_{0,2}$ line is the strongest; the 3$_{2,1}-$2$_{2,0}$ and 3$_{2,2}-$2$_{2,1}$ lines are weaker by a factor of 2 or more.
The spectra of the 3$_{2,1}-$2$_{2,0}$ and 3$_{2,2}-$2$_{2,1}$ lines only have two velocity channels.
For the 3$_{2,2}-$2$_{2,1}$ line at P2, the spectrum was not well fitted because of the peak intensity lower than 3$\sigma$ noise level and the narrow velocity width.
At P4, the spectrum of the 3$_{2,2}-$2$_{2,1}$ line shows absorption at $\sim 6$ \kms or asymmetric line profile.
Using these spectra except for the 3$_{2,2}-$2$_{2,1}$ line at P2, we derive the physical parameters at these positions in the next section.

\begin{table}[htbp!]
  \caption{Line Parameters at the four positions in the \hhco\ map$^a$ \label{line_para}}
  \centering
  \begin{tabular}{ccccc}
  \hline
  Position& Transition & $T_{\rm peak}$ &  $V_{\rm LSR}$   & FWHM  \\  %
  && (K)&(\kms)& (\kms) \\\hline
\hline
P1 & 3$_{0,3}-$2$_{0,2}$&9.65 (0.35)  &    4.40 (0.03)&  1.47 (0.06)\\
& 3$_{2,1}-$2$_{2,0}$ &1.92 (0.49) &    4.64 (0.24)&  1.92 (0.57)\\
& 3$_{2,2}-$2$_{2,1}$ &2.32 (0.51)&  4.35 (0.18)&  1.66 (0.42)\\
\hline
P2&  3$_{0,3}-$2$_{0,2}$ &8.18 (0.34)&  4.8 (0.02)&  1.24 (0.06)\\
& 3$_{2,1}-$2$_{2,0}$ &2.89 (0.49)&  5.06 (0.08)&  0.81 (0.17)\\
& 3$_{2,2}-$2$_{2,1}$ &  - &  - & - \\
\hline
P3& 3$_{0,3}-$2$_{0,2}$ &4.28 (0.19)&  4.32 (0.05)&  2.25 (0.12)\\
& 3$_{2,1}-$2$_{2,0}$ &1.46 (0.20)&  4.61 (0.14)&  2.12 (0.33)\\
& 3$_{2,2}-$2$_{2,1}$ &1.31 (0.21)&  4.52 (0.15)&  1.89 (0.35)\\
\hline
P4 & 3$_{0,3}-$2$_{0,2}$ & 4.22 (0.30)&  4.3 (0.15)&  2.05 (0.34)\\
& 3$_{2,1}-$2$_{2,0}$ &1.34 (0.22)&  4.34 (0.28)&  1.75 (0.65)\\
& 3$_{2,2}-$2$_{2,1}$ &1.66 (1.24)&  5.49 (3.27)&  4.61 (3.73)\\
  \hline
  \end{tabular}
  \begin{flushleft}
  \tablecomments{$^a$ Each position is shown in Figure \ref{spectra_hhco}. The parentheses show the uncertainties of corresponding quantities.}
  \end{flushleft}
\end{table}

\subsection{Physical Parameters of \hhco\ Derived from Non-LTE Analyses}
\label{sec:non-lte-hhco}

The gas kinetic temperature for each position (P1, P2, P3, and P4) is evaluated from the \hhco\ lines using a non-LTE (local
thermodynamic equilibrium) method.
We use the fitting results in Table \ref{line_para}.
Although we have three lines, the upper state energies of 3$_{2,1}-$2$_{2,0}$ and 3$_{2,2}-$2$_{2,1}$ are too close in the upper-state energy to determine the H$_2$ density as well as the kinetic temperature and the column density of \hhco.
Hence, we derive the kinetic temperature and the column density of \hhco\ by assuming three different values for the H$_2$ density (10$^5$, 10$^6$, and 10$^7$ cm$^{-3}$).
The results are summarized in Table \ref{hhco_temp}.

The temperatures have a large uncertainty which is derived from the $\chi^2$ analysis. 
At P1, the upper limit of the temperature is determined to be 37 K for the H$_2$ density of 10$^5$ cm$^{-3}$, while the temperature is 50$^{+16}_{-11}$ K for the H$_2$ density of 10$^6$ cm$^{-3}$.
Except for these cases, the lower limits of the temperatures are similar for each H$_2$ density, 40-52 K, 75-80 K, and 65-75 K for the H$_2$ density of 10$^5$, 10$^6$, and 10$^7$ cm$^{-3}$, respectively.
These kinetic temperatures are much lower than the excitation temperature (rotation temperature) of H$_2$ that we derive in Section \ref{sec:TempH2}, supporting that these lines would trace the physical structures with different temperatures.
 



\begin{table}[htbp!]
\caption{Temperatures and Column Densities of \hhco$^a$ \label{hhco_temp}}
\centering
\begin{tabular}{ccccccc}
\hline
\hline

 Position$^b$ & 10$^{5}$  & 10$^{6}$  & 10$^{7}$ \\
          &(cm$^{-3}$)&(cm$^{-3}$)&(cm$^{-3}$)\\
\hline
          & \multicolumn{3}{c}{$T$ (K)}\\
\hline
P1 & $<$37            & 50$^{+16}_{-11}$   & 91$^{+54}_{-21}$  \\
P2 & 64$^{+32}_{-20}$ & 120$^{+105}_{-40}$ & 110$^{+75}_{-35}$  \\
P3 & 70$^{+28}_{-18}$ & 110$^{+66}_{-33}$  & 90$^{+54}_{-21}$ \\
P4 & 76$^{+48}_{-24}$ & 120$^{+140}_{-45}$ & 100$^{+105}_{-35}$  \\
\hline
           & \multicolumn{3}{c}{$N$ (para) (10$^{14}$ cm$^{-2})$}\\
\hline
P1 &  4.4$^{+6.5}_{-1.5}$ & 0.62$^{+0.08}_{-0.06}$ & 0.6$^{+0.1}_{-0.1}$ \\
P2 &  0.8$^{+0.2}_{-0.1}$ & 0.26$^{+0.03}_{-0.03}$ & 0.5$^{+0.1}_{-0.1}$ \\
P3 &  0.9$^{+0.2}_{-0.1}$ & 0.38$^{+0.04}_{-0.04}$ & 0.6$^{+0.1}_{-0.1}$ \\
P4  & 1.2$^{+0.2}_{-0.2}$ & 0.52$^{+0.09}_{-0.07}$ & 0.9$^{+0.4}_{-0.2}$ \\

\hline
\end{tabular}
\begin{flushleft}
  \tablecomments{
$^a$ The H$_2$ density is assumed from 10$^{5}$ to 10$^{7}$ cm$^{-3}$. 
The ortho to para ratio of \hhco\ is assumed to be 3.
$^b$ Each position is shown in Figure \ref{spectra_hhco}. The uncertainties are estimated by using a $\chi^2$ analysis.}
\end{flushleft}
\end{table}

\end{document}